\documentclass{aa}
\usepackage[english]{babel}
\usepackage[varg]{txfonts}
\usepackage{graphicx}
\usepackage{hyperref}

\begin{document}

\title{Observations and modeling of spectral line asymmetries in stellar flares}

\author{J. Wollmann \inst{1, 2} \and P. Heinzel\inst{1, 3} \and P. Kabáth\inst{1}}

\institute{Astronomical Institute of the Czech Academy of Sciences, Fri\v{c}ova 298, CZ-25165 Ond\v{r}ejov, Czech Republic \and Charles University, Astronomical Institute, V Hole\v{s}ovi\v{c}k\'ach CZ-18000, Praha 8, Czech Republic \and Center of Scientific Excellence - Solar and Stellar Activity, University of Wroc\l aw, Kopernka 11, 51-622 Wroc\l aw, Poland}

\date{}

\abstract
   {Stellar flares are energetic events occurring in stellar atmospheres. They have been observed on various stars using photometric light curves and spectra. On some cool stars, flares tend to release substantially more energy compared to solar flares. Spectroscopic observations have revealed that some spectral lines, aside from an enhancement and broadening, exhibit asymmetry in their profile. Asymmetries with enhanced blue wings are often associated with the presence of coronal mass ejections while the origin of the red asymmetries is currently not well understood. A few mechanisms have been suggested but no modeling has been performed yet.}
   {We observed the dMe star AD Leo using the 2-meter Perek telescope at Ondřejov observatory, with simultaneous photometric light curves. In analogy with solar flares, we model the H$\alpha$ line emergent from an extensive arcade of cool flare loops and explain the observed asymmetries using the concept of coronal rain.}
   {We solve the non-LTE radiative transfer in H$\alpha$ within cool flare loops taking into account the velocity distribution of individual rain clouds. For a flare occurring at the center of the stellar disc, we then integrate radiation emergent from the whole arcade to get the flux from the loop area.}
   {We observed two flares in the H$\alpha$ line that exhibit red wing asymmetry corresponding to velocities up to 50\,km s$^{-1}$ during the gradual phase of the flare. Synthetic profiles generated from the model of coronal rain have enhanced red wings quite compatible with observations.} 
   {}
   

\maketitle

\bibpunct{(}{)}{;}{a}{}{,} 

\section{Introduction}
Stellar flares are energetic events occurring due to the energy release during magnetic reconnection in the atmospheres of certain stars. Based on analogy with the Sun it is commonly assumed that they are stellar counterparts of solar flares. However, contrary to solar flares, we cannot spatially resolve stellar flaring structures and thus we are tempted to use our global picture of solar flares to understand stellar flares. This is particularly true when we consider different geometrical
and physical conditions in chromospheric ribbons and extended hot or cool flare loops.
Stellar flares were observed on a variety of spectral types of stars \citep{Pettersen1989} but a majority of them occur on cool dMe type stars. The energy released during flares on such stars is typically of the order of 10$^{31}$ -- 10$^{34}$ ergs, but can reach up to 10$^{36}$ ergs or even more in case of the so-called superflares \citep{Shibata2016}. That is about four orders of magnitude more as compared to the largest solar flares with the energy of 10$^{32}$ ergs. The
reason is that dMe stars are expected to have stronger and extended magnetic fields compared to the
Sun \citep{Crespo2015}.

Most observations of stellar flares are photometric ones. These observations may have a high temporal resolution even in the case of rather faint dMe stars (which have low effective temperatures) and therefore are useful for the determination of the flare occurrence and estimation of some flare properties like the total energy released, for example \citep{Pietras2022}, \citep{Doyle2019} and \citep{Medina2020}.
However, well-resolved spectral observations must be used to study stellar flare dynamics, via the detection of Doppler shifts or line asymmetries.
Spectroscopic studies have revealed significant changes in the spectra during a flare. Aside from changes in the continuum and spectral line strengths, various chromospheric lines, namely hydrogen Balmer lines, exhibit significant broadening and profile asymmetry. The latter is well known as the so-called blue or red asymmetry with enhanced wing intensities. These are usually assigned to dynamics of the chromospheric condensation or evaporative processes, both in solar and analogical stellar cases. Blue wing enhancements can be associated with the presence of coronal mass ejections (CME), see e.g. \citet{Muheki2020cme}, \citet{Vida2019} and \citet{Leitzinger2022}. 
Recently \citet{Muheki2020} showed time evolution of the hydrogen H$\alpha$ line asymmetry where the red wing of the emission line is enhanced in the case of AD Leo dMe star. A similar enhancement was also detected by \citet{Wu2022} in various lines including H$\alpha$. In both papers the authors suggest several possibilities to explain such asymmetries, but no modeling was performed so far. An important observation is that the red-wing enhancement appears at wavelength positions that correspond to rather large Doppler velocities exceeding 100~km\,s$^{-1}$ which, in the case of solar flares, was never detected during the gradual phase. In this paper, we present similar H$\alpha$ line observations of the star AD Leo conducted with the Ond\v{r}ejov Echelle Spectrograph (OES) attached to the 2-meter Perek telescope. Because of large Doppler shifts during the gradual phase, we suggest here that such red asymmetries are caused by downward flows of cool plasma blobs along extended flare loops. Such phenomenon is well known in the case of solar flares where it was called loop prominences, post-flare loops (but see \citep{Svestka2007}) or recently 'a coronal rain' \citep{Antolin2010}.
In order to quantitatively reproduce our observations, we develop an approximate non-LTE radiative-transfer model to synthesize H$\alpha$ line profiles from the spatially unresolved extended arcade of cool flare loops and we compare the results of our simulations with OES observations.

\section{Observations}
\label{sec:observations}
Stellar flares can be observed via photometric light curves especially using U and B filters. The typical shape of these light curves during a flare is a sharp rise of the flux at the beginning of the flare followed by a gradual fall to the preflare level. Flares can also be observed using the spectra of the stars, for example using a light curve of an integrated flux of some spectral lines and continua.

\subsection{Dataset}
For our study, we observed the dMe star AD Leo in three periods (observing campaigns) during spring 2019, 2020, and 2021 using the 2-meter Perek telescope at Ondřejov observatory (Czech Republic). Spectra were obtained using Ondřejov Echelle Spectrograph (OES), a broad range spectrograph with range 4250\,Å--7500\,Å for 2019 and 2020 observations and 3900\,Å--9000\,Å for 2021. The resolving power of the spectrograph is $R$~$\sim$~40000 in the H$\alpha$ region. Exposure times were 15\,minutes for 2019 and 2020 observations and 10\,minutes for 2021 to get an acceptable signal-to-noise ratio. With these exposure times we are getting at least 9-10 signal-to-noise ratio in the H$\alpha$ region in the center of the echelle order. Other properties of the OES and Perek telescope are described by \citet{Kabath2020}. Spectra were extracted and bias-flat corrected using standard IRAF procedures.

During the 2019 and 2021 campaigns, astronomers from the SPHE (section of variable stars and exoplanets) group of the Czech Astronomical Society simultaneously observed AD Leo using different photometric filters from various locations in the Czech Republic. Reduced light curves by the observers were provided to us with typical exposure times ranging from 30--90\,s.

\subsection{Line profiles}
In order to quantitatively study changes of spectral line profiles during flares, we need to calibrate the observed spectra to absolute radiometric units. For that purpose, our calibration method requires a quiescent calibrated spectrum of AD Leo. We observed a spectrum of ESO spectrophotometric standard star HD~93521 \citep{Oke1990} during the campaign in 2020 and just after that a spectrum of AD Leo. From the ratio of the observed spectra and known calibrated spectrum of HD~93521 we get calibrated quiescent AD Leo spectrum $F_{\lambda}^{quiescent}$. We used light curves to ensure that there were no variations (e.g. flaring activity) during exposure of the quiescent spectrum.

Since the Earth's atmospheric extinction can vary on timescales of flares and we also expect that continuum flux changes during the course of a flare, we needed to account for these variations during our calibration process. For calibration of spectra during the flare, we assumed that the spectra observed before (at time $t_1$) and after (at time $t_2$) a single flare correspond to the calibrated quiescent spectrum we observed before. To determine these we used the photometric light curves. Under this assumption, we apply linear interpolation to calculate expected quiescent spectra during the flare between just before and after the flare spectra. From the ratio of the observed spectrum during the flare and calculated expected quiescent spectrum during the flare we obtain spectra calibrated to absolute units during the flare. The expression is following:
\begin{equation}
    F_{\lambda}^{flare}(t) = \frac{f_{\lambda}(t)}{f_{\lambda}(t_1) + \frac{f_{\lambda}(t_2) - f_{\lambda}(t_1)}{t_2 - t_1} \left(t - t_1 \right)} F^{quiescent}_{\lambda}
\end{equation}
where $f_{\lambda}(t)$ is uncalibrated flux from telescope during the flare, $f_{\lambda}(t_1)$ is uncalibrated flux prior to the flare, $f_{\lambda}(t_2)$ is uncalibrated flux after the flare and $F_{\lambda}^{quiescent}$ is calibrated quiescent flux.

During the flare, the continuum is varying. In the case of the Sun, this was demonstrated many times, see \citet{Kuhar2016}. In the case of stellar flares, the visible continuum seems to be enhanced much more compared to the Sun and in the case of strong flares can be fitted by a Planck function at temperatures around 10000\,K \citep{Kowalski2013}. However, all these observations typically refer to flare ribbons without any discussion of the possible contribution of the flare loops to the continuum. The only such study was performed by \citet{Heinzel2018} who showed that the arcade of loops may contribute significantly to the total stellar continuum enhancement. In the present study, we do not focus on the analysis of the nature of the continuum variations which may be caused by both the loops and ribbons. The continuum is certainly enhanced during our observations which is manifested by the optical light curves. In the observed flare spectra, the continuum variations are present due to the flare itself and the atmospheric transmission effects. To eliminate the latter, we interpolate between the before-the-flare and the after-the-flare "quiescent" spectra. Certainly, this linear interpolation does not contain short-term atmospheric fluctuations which can introduce uncertainties that affect the observed continuum levels.

\begin{figure}
    \centering
    \resizebox{\hsize}{!}{\includegraphics{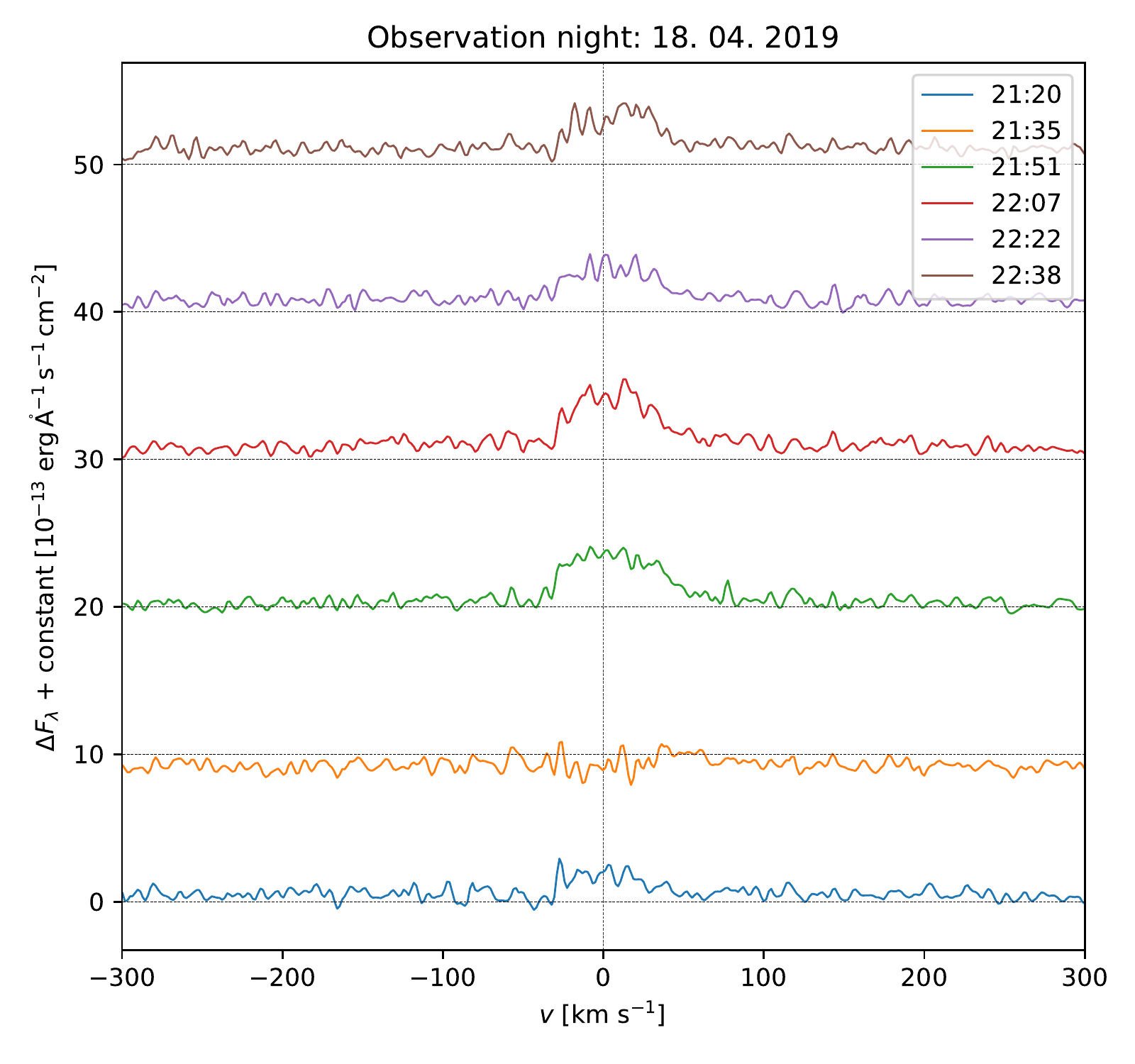}}
    \caption{Difference between H$\alpha$ profile during the flare and the quiescent observed on AD Leo. We added a constant (dashed line) to every spectrum increasing with time. We can see that the H$\alpha$ line flux is enhanced and the profile is asymmetric with the enhanced red wing of the line.}
    \label{fig:profile_changes_2}
    \centering
    \resizebox{\hsize}{!}{\includegraphics{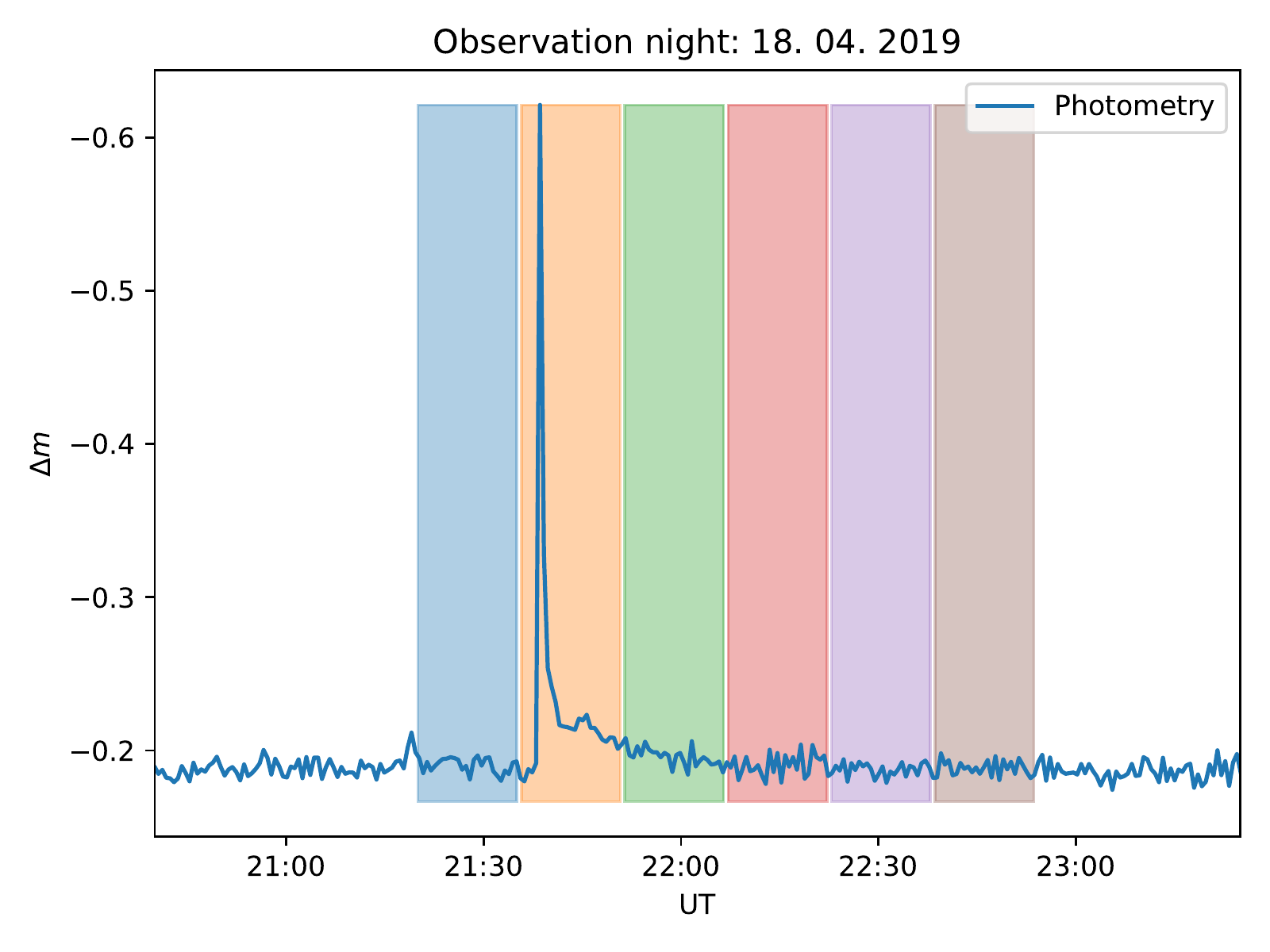}}
    \caption{Photometric light curve of AD Leo flare with marked exposure intervals for spectra shown in Fig. \ref{fig:profile_changes_2}. The light curve is a relative difference in the magnitude in the B filter.}
    \label{fig:profiles_with_photometry_2}
\end{figure}

\begin{figure}
    \centering
    \resizebox{\hsize}{!}{\includegraphics{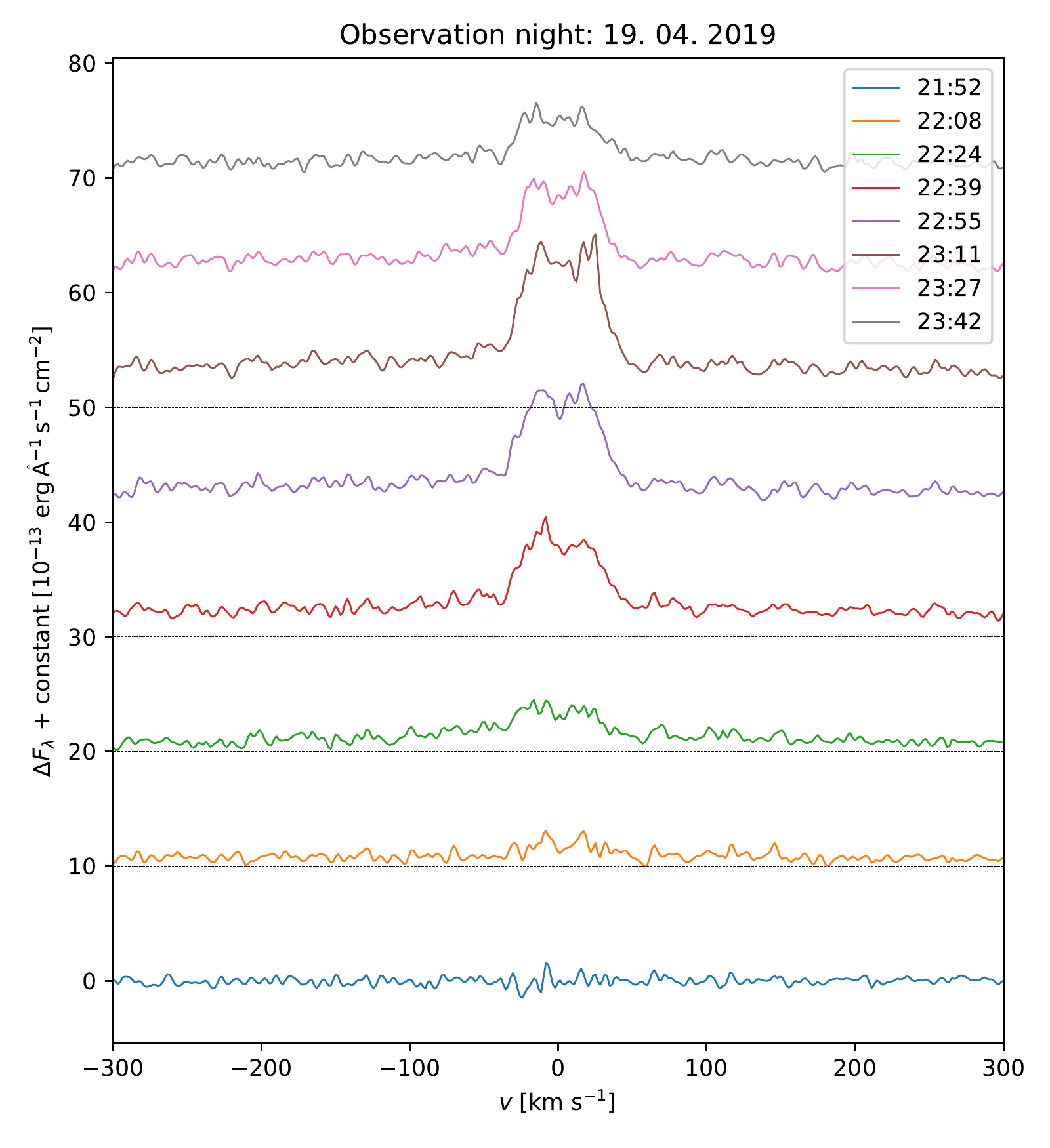}}
    \caption{Difference between H$\alpha$ profile during the flare and the quiescent observed on AD Leo. We added a constant (dashed line) to every spectrum increasing with time. We can see that the H$\alpha$ line flux is enhanced and the profile is asymmetric with the enhanced red wing of the line.}
    \label{fig:profile_changes_1}
    \centering
    \resizebox{\hsize}{!}{\includegraphics{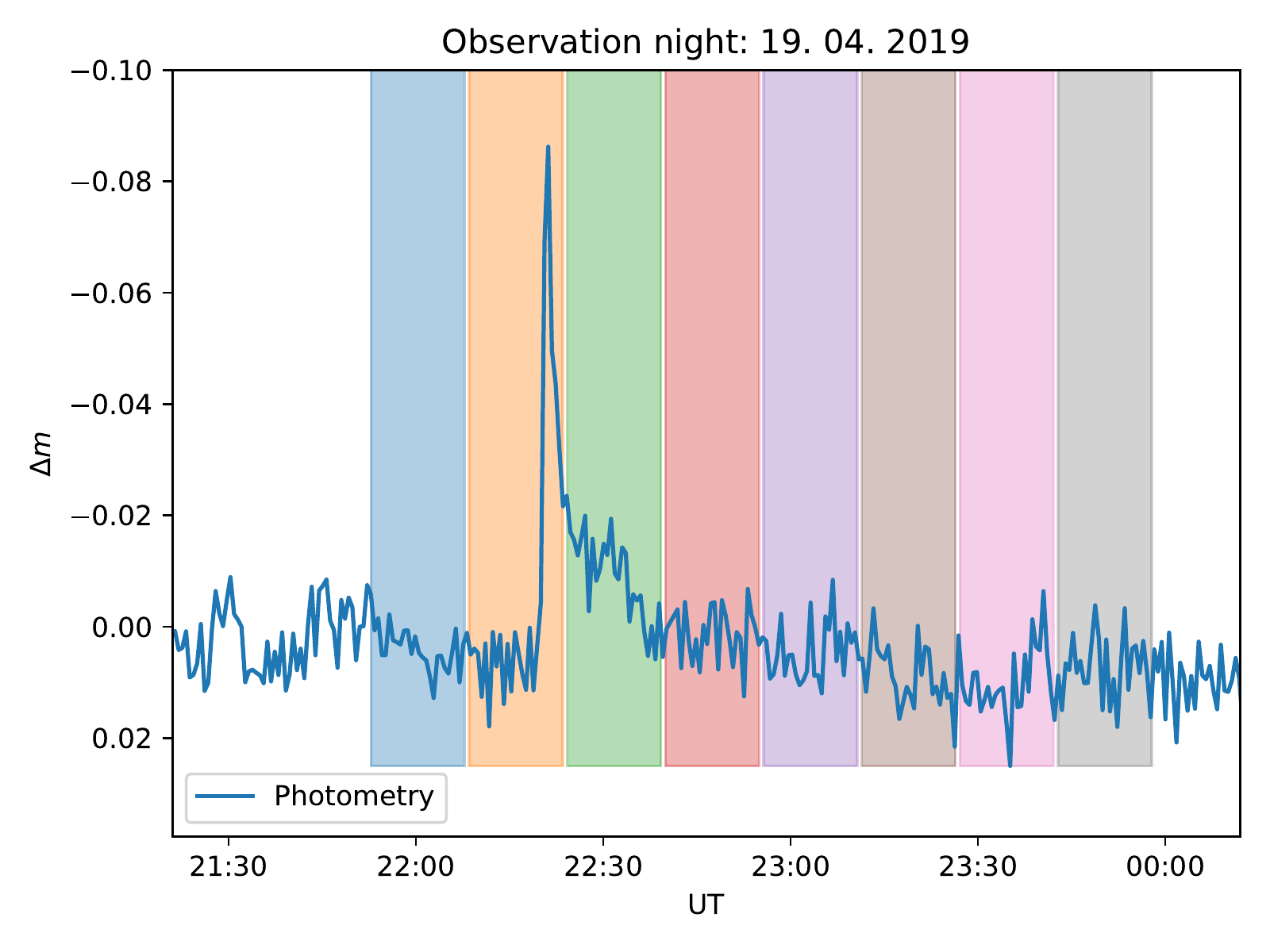}}
    \caption{Photometric light curve of AD Leo flare with marked exposure intervals for spectra shown in Fig. \ref{fig:profile_changes_1}. The light curve is a relative difference in the magnitude in the B filter.}
    \label{fig:profiles_with_photometry_1}
\end{figure}

We study the effect of flares on the H$\alpha$ line. To show this effect we plot $\Delta F_{\lambda}$, the flux difference between calibrated line profile during the flare and the quiescent calibrated profile:
\begin{equation}
    \Delta F_{\lambda}(t) = F_{\lambda}^{\rm{flare}}(t) - F_{\lambda}^{\rm{quiescent}}
    \label{flux_difference}
\end{equation}
where $F_{\lambda}^{\rm{flare}}(t)$ is calibrated flux observed during a flare and $F_{\lambda}^{\rm{quiescent}}$ is quiescent calibrated flux. We observed two flares with good spectra coverage that were not disturbed by other flares and used our calibration method. Results are shown in the Fig. \ref{fig:profile_changes_2} and \ref{fig:profile_changes_1}. Additionally to show the correlation between changes of the profile of the H$\alpha$ line and the photometric light curve in the B filter we plot spectra exposure times along with the light curve in the Fig. \ref{fig:profiles_with_photometry_2} and \ref{fig:profiles_with_photometry_1}. Color blocks in Fig. \ref{fig:profiles_with_photometry_2} and \ref{fig:profiles_with_photometry_1} correspond to the exposure time of the spectra in the Fig. \ref{fig:profile_changes_2} and \ref{fig:profile_changes_1} respectively.

The light curve in Fig. \ref{fig:profiles_with_photometry_2} shows that the first spectrum was exposed during the preflare phase and the next spectrum during the impulsive phase and partially during the gradual phase. The H$\alpha$ line during the preflare phase shows a little rise in the line center, see Fig. \ref{fig:profile_changes_2}. That could be caused by a smaller-scale flare (the small bump in the light curve) occurring at about 21:20 UT just prior to the exposure of the first spectrum. The next spectrum shows a very small decrease in continuum and the line compared to the quiescent flux. This could be caused by additional uncertainties introduced by linear interpolation used for calibration. During later stages, the continuum does not appear to change much. In the later stages of the gradual phase, we can see that the H$\alpha$ profile exhibits enhancement in the line center and in the wings. During this flare, the red wing (positive velocity) is stronger compared to the blue wing (negative velocity) within the range of 15\,--\,50~km\,s$^{-1}$ creating an asymmetrical profile.

The light curve of the second flare in Fig. \ref{fig:profiles_with_photometry_1} shows that the first calibrated spectrum was exposed during the preflare phase. The corresponding H$\alpha$ profile shows no change in Fig. \ref{fig:profile_changes_1}. Next exposure captured the spectrum during the impulsive phase of a flare (the sharp rise in magnitude) but the H$\alpha$ profile only exhibits a little rise in flux. The next (green) spectrum shows a little enhancement in the line center and also in the blue part of the wing creating a small blue wing asymmetry at velocities 15\,--\,30~km\,s$^{-1}$. Following spectra were observed during the gradual phase (slow fall of flux to preflare levels in light curves). During this phase, the flux in the H$\alpha$ line begins to increase compared to the quiescent state. One can observe that during this phase the red wing (positive velocity) of the line is slightly higher than the corresponding velocity in the blue wing (negative velocity) within the range of 15\,--\,50~km\,s$^{-1}$ creating an asymmetrical profile. Additionally the continuum exhibits an enhancement during the gradual phase.

\section{Model}

According to the standard model of solar flares, see e.g. \citet{Priest2014}, the energy released during the magnetic reconnection creates a flux of high energetic particles (electrons) that travel along the reconnected magnetic field lines to lower layers of the solar atmospheres. In the region of higher density, in the chromosphere, these particles transfer their energy to the surrounding plasma by heating it. These regions are bright sources of hard X-ray radiation and are known as flare ribbons.
Heated plasma in the chromosphere is pushed down forming the so-called chromospheric condensation (with velocities of a few tens of km s$^{-1}$) and simultaneously the upper chromosphere and low transition region evaporate into the hot loop. These hot loops then cool down and form cool loops visible in the H$\alpha$ line and many other lines of different species e.g. \citet{Mikula2017}. Downflows in the cool loops have much larger velocities than in the chromospheric condensation and this looks consistent also with stellar observations.
This coronal rain is prominent in H$\alpha$ and high-resolution images and a movie can be seen for example in \citet{Jing2016}. We use this example of an extended flare-loop arcade as a prototype of a stellar case.

\begin{figure}
    \centering
    \resizebox{\hsize}{!}{\includegraphics{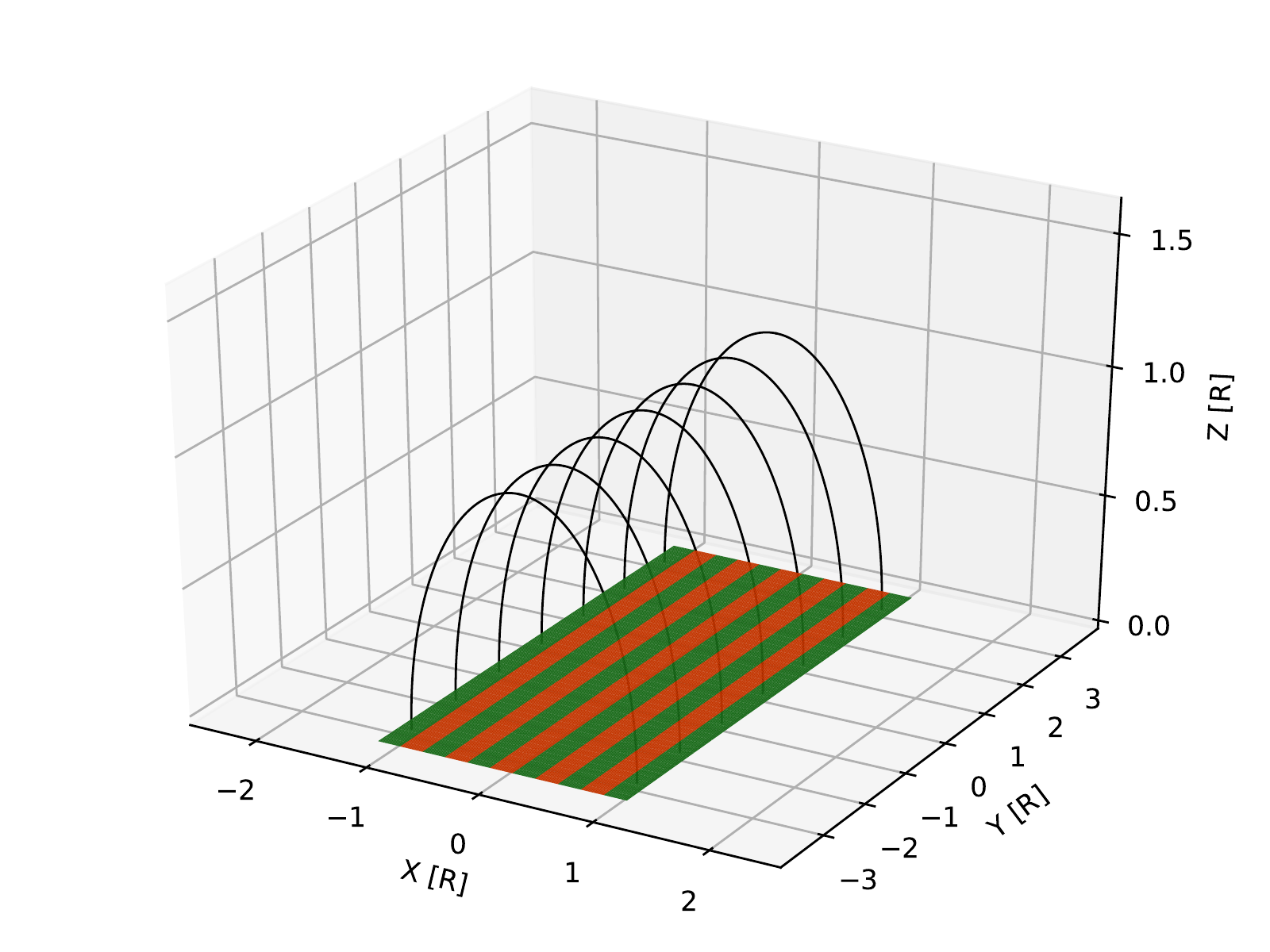}}
    \caption{Flare loop arcade structure. Semicircular field lines are shown in black. The green and red stripes mark the stellar surface and plasma projected onto the stellar surface in a single stripe is assumed to have the same velocity. Coordinates are expressed in the radius of the semicircular field lines.}
    \label{fig:archade_scheme}
\end{figure}

In our model, we solve the non-LTE radiative transfer in the H$\alpha$ line within flare loops that form an arcade-like structure and we only model a situation when flare loops are already formed. The model is inspired by the presence of a coronal rain that occurs on the Sun during solar flares. Our model is based on two assumptions about the general properties of the flare. Flare occurs in the center of the stellar disk with respect to the observer and the plasma in flare loops is moving along the semicircular magnetic field lines in a free fall. The arcade structure scheme is shown in Fig. \ref{fig:archade_scheme}.

\begin{figure}
    \centering
    \resizebox{\hsize}{!}{\includegraphics{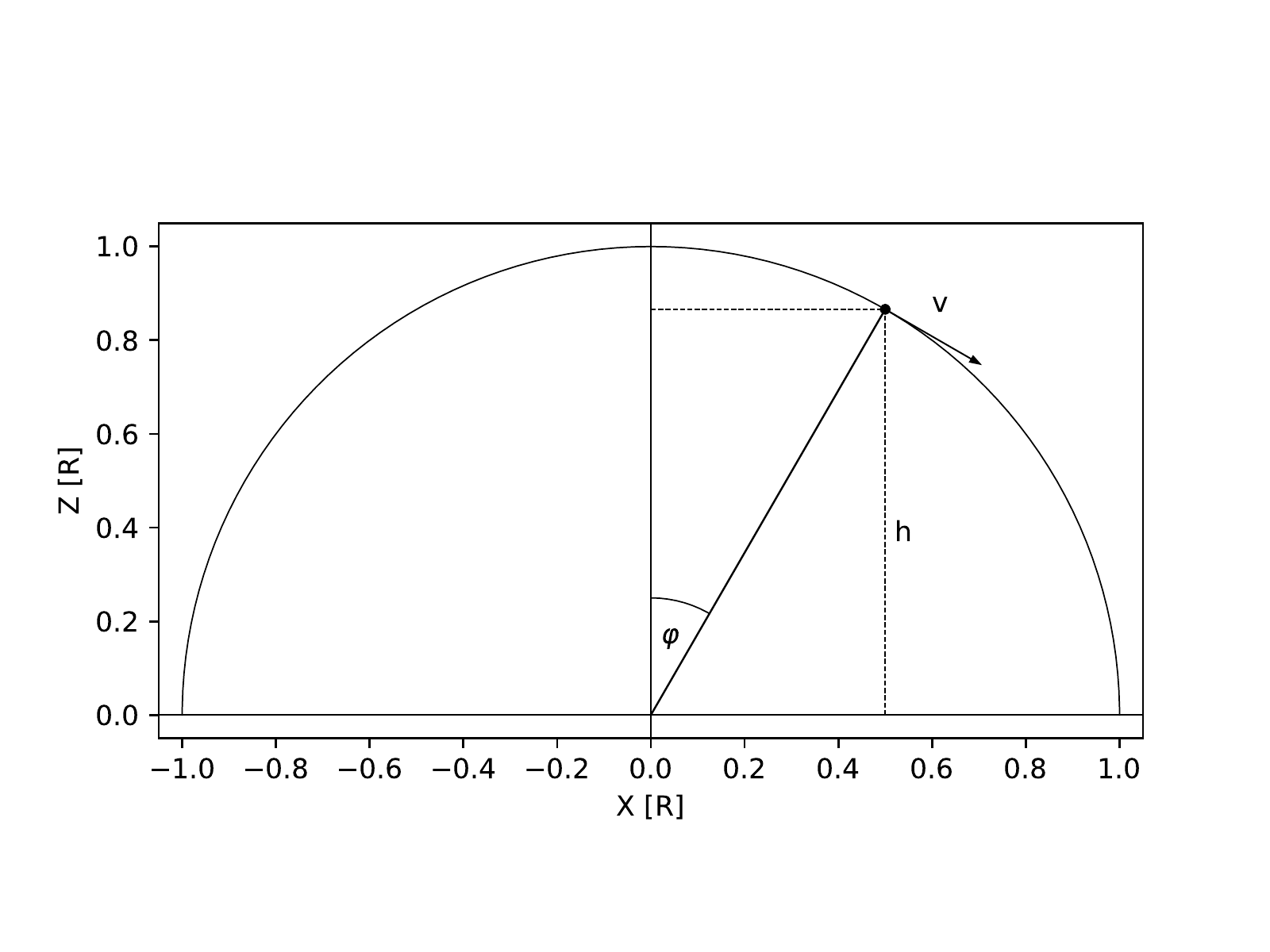}}
    \caption{Scheme for free fall solution of a cloud of plasma falling along the semicircular magnetic field line. Plane Z = 0 represents stellar surface.}
    \label{fig:free_fall_scheme}
\end{figure}

To solve the radiation transfer through the flare loops we use the cloud model of plasma (plasma is structured into smaller clouds) with the same physical properties \citep{Tziotziou2007}. These clouds are moving in free fall along the magnetic field lines towards the stellar surface with initial velocity v$_0$. Given this setup, the clouds with the same X coordinate in the Fig. \ref{fig:archade_scheme} (along the arcade) have the same velocity. This is indicated in the Fig. \ref{fig:archade_scheme}. The clouds projected onto the stellar surface in the same stripe have the same velocity.
Velocity for a given X position can be obtained by solving Lagrange equations for Lagrangian:
\begin{equation}
    L = \frac{1}{2} m r^2 \dot{\varphi}^2 - m g r \cos (\varphi)
\end{equation}
where $g$ is surface gravity acceleration of the star, $r$ is the radius of the flare loop along which the cloud is moving and $\varphi$ is angle describing cloud position in Fig. \ref{fig:free_fall_scheme}. We used explicit 5th order Runge-Kutta method to solve the equation with the initial condition $\varphi (0) = 0$ and $\dot{\varphi}(0) = \frac{v_0}{r}$.

Since the clouds have the same physical properties except for relative velocity towards the stellar surface we can calculate the spectrum of the arcade as a sum of radiation over the flare area:
\begin{equation}
    I^{\rm{flare}}(\nu, \tau=0) = \frac{\sum^{n}_{k=1} I^{\rm{cloud}} (\nu, \tau=0, v_{k}) A_{k}}{A^{\rm{flare}}}
\label{arcade_intensity}
\end{equation}
where $I^{\rm{cloud}}$ is the specific intensity of a cloud seen by a static observer with respect to the stellar surface, $\nu$ is the frequency, $\tau$ is the optical thickness of clouds of plasma, $v_{k}$ is the vertical component of the velocity of all clouds in a k-th stripe, $A_{k}$ is the area of a k-th stripe (area composed of clouds with the same velocity and same height) and $A^{\rm{flare}}$ is the total area of a flare. In our case it is the area of the whole extended arcade as is discussed by \citet{Heinzel2018}. The radiation coming from clouds must be correctly Doppler-shifted.

\subsection{Intensity of a single static cloud}
The specific intensity of radiation from a single cloud is given as the solution of the radiative transfer equation
\begin{equation}
    \frac{dI}{d \tau}(\nu, \tau) = S(\nu, \tau) - I(\nu, \tau)
\end{equation}
where $S$ is the source function. The formal solution of the equation is a combination of two terms: the background disc radiation attenuated by the plasma and the source term of the cloud. The exact formula is:
\begin{equation}
    I(\nu, \tau = 0) = I (\nu, \tau_0(\nu)) e^{-\tau_0(\nu)} + \int^{\tau_0} _0 S(t(\nu)) e^{-t(\nu)} dt(\nu)
    \label{general_solution_radiative_transfer}
\end{equation}
where $\tau_0(\nu)$ is the optical thickness of clouds at a given frequency. Assuming optical depth at the line center $\tau_0$ we can express $\tau_0(\nu) = \tau_0 \phi(\nu)$ where $\phi(\nu)$ is line profile:
\begin{equation}
    \phi(\nu) = \frac{1}{\sqrt{\pi \Delta} \nu} e^{-\left(\frac{(\nu - \nu_0)}{\Delta \nu}\right)^2}
\end{equation}
where $\nu_0$ is H$\alpha$ line-center frequency and $\Delta \nu$ is:
\begin{equation}
    \Delta \nu = \frac{\nu_0}{c} \sqrt{\frac{2 k_B T}{m} + v_t^2}
\end{equation}
where $T$ is plasma temperature, $k_B$ is Boltzmann constant, $m$ is hydrogen mass and $v_t$ is the turbulent velocity of plasma.
We assume that the background radiation incoming to clouds, $I(\nu, \tau_0(\nu))$, is equal to the stellar quiescent radiation without flare.

We thus need to determine the source function S. For the H$\alpha$ line we use here a two-level atom approximation which allows us to write the source function as a sum of scattering term and thermal terms (e.g. \citet{Heinzel2019})
\begin{equation}
    S = \left( 1 - \epsilon \right) \bar{J} + \epsilon B_{\nu_0}
    \label{source_function}
\end{equation}
where $\epsilon$ is photon destruction probability, $B_{\nu_0}$ is Planck's function at the line center frequency with the temperature of the plasma in clouds $T$ and $\bar{J}$ is the mean integrated intensity of the radiation field inside the cloud. Following \citet{Rybicki1984} we can split $\bar{J}$ as $\bar{J} = \bar{J}^{\rm dif} + \bar{J}^{\rm dir}$, where $\bar{J}^{\rm dif}$ is the diffuse part of the intensity and $\bar{J}^{\rm dir}$ is the part due to direct external illumination of the cloud. Here we present an approximate formula for $S$ which was obtained in a heuristic way by Heinzel \& Wollmann (2022, in preparation). Writing formally
\begin{equation}
    S^{\rm dir} = \left( 1 - \epsilon \right) \bar{J}^{\rm dir} + \epsilon B_{\nu_0}
    \label{source_function_with_J_dir}
\end{equation}
we get 
\begin{equation}
    S(\tau_0/2) = 
    \frac{S^{\rm dir}(\tau_0/2)}{1 - (1-\epsilon)[1-K_2(\tau_0/2)]} \, ,
\end{equation}
where $K_2$ is the function evaluated numerically according to {\citet{Hummer1982}.
Its value goes to one for $\tau$ approaching zero and decreases with increasing $\tau$. We thus see that for very small $\tau$ S approaches $S^{\rm dir}$ as expected.
For optically-thin clouds, $\bar{J}^{\rm dir}$ is evaluated directly as 
\begin{equation}
    \bar{J}^{\rm dir} = \int^{\infty}_0 J^{\rm inc}(\nu) \phi(\nu) d\nu
    \label{j_dir_simple}
\end{equation}
where $\phi(\nu)$ is the absorption line profile. We can obtain the mean incident radiation intensity simply using the dilution factor $W$ (e. g. \citep{Jejcic2009})
\begin{equation}
    J^{\rm inc}(\nu) = W I(\nu) \, ,
\end{equation}
where $I(\nu)$ is the stellar quiescent specific intensity and $W$ is
\begin{equation}
    W = \frac{1}{2} \left[ 1 - \left(1 - \frac{R^2}{\left(R + H\right)^2} \right)^{\frac{1}{2}} \right] \, .
\end{equation}
$R$ is the stellar radius and $H$ is the height above the stellar surface.
However, for general values of $\tau$ we multiply this direct mean intensity by a correction factor as follows
\begin{equation}
    \bar{J}^{\rm dir} = \int^{\infty}_0 J^{\rm inc}(\nu) \phi(\nu) d\nu \times
    (1 - {\rm e}^{-\tau_0/2})/(\tau_0/2) \, .
    \label{j_dir_general}
\end{equation}
Note that this factor goes to one for 
$\tau << 1$ and we get an optically-thin value as in Eq. \ref{j_dir_simple}.
The above-described approach gives a good estimate of the
H$\alpha$ line source function in externally illuminated clouds, to within a factor of two. For details and comparisons with exact
solutions see Heinzel \& Wollmann (2022, in preparation).

The photon destruction probability $\epsilon$ is:
\begin{align}
    \epsilon &= \frac{\epsilon'}{1 + \epsilon'}\\
    \epsilon' &= \frac{C_{32}}{A_{32}} \left(1 - \rm{e}^{- \frac{h \nu_0}{k_B T}} \right)
\end{align}
where $A_{32}$ is Einstein coefficient of spontaneous emission and $C_{32}$ is the collision coefficient which can be approximated according to \citet{Johnson1972} as:
\begin{equation}
    C_{32} = n_e f(T)
    \label{c_21}
\end{equation}
where $n_e$ is the electron density in the clouds and $f(T)$ is a weak function of temperature. Values of $f(T)$ for selected temperature values are listed in Table \ref{temperatue_collisional_rates_depencency}.

\begin{table}
    \caption{Value of $f(T)$ for a few selected temperature values.}
    \label{temperatue_collisional_rates_depencency}
    \centering
    \begin{tabular}{c c}
         \hline \hline
         $T$ [K] & $f(T)$ [cm$^{3}$ s$^{-1}$] \\
         \hline
         8000 & 2.501 $\times 10^{-7}$ \\
         10000 & 2.529 $\times 10^{-7}$ \\
         15000 & 2.649 $\times 10^{-7}$ \\
         20000 & 2.777 $\times 10^{-7}$ \\
         \hline
    \end{tabular}
\end{table}

Since the source function in our approximation is constant inside the cloud,
Eq. \ref{general_solution_radiative_transfer} simplifies to:
\begin{equation}
    I(\nu, \tau=0) = I (\nu, \tau_0(\nu)) \rm{e}^{-\tau_0(\nu)} + S \left[ 1 - \rm{e}^{-\tau_0(\nu)} \right]
    \label{final_intensity}
\end{equation}

\subsection{Intensity of a single moving cloud}
When the cloud is moving with respect to the stellar surface we need to account for Doppler shifts by modifying both terms in the equation \ref{final_intensity}. We calculate the intensity coming from the cloud with the vertical component of velocity $v$ towards the stellar surface in the reference frame connected to the stellar surface.

Background radiation in the first term is already in the correct frame so it does not shift but it is attenuated by the plasma in the cloud which "sees" the radiation at shifted (higher) frequency.
\begin{equation}
I^{\rm{attenuated}}(\nu, \tau=0, v) = I(\nu, \tau_0(\nu)) \rm{e}^{-\tau_0(\nu^+)}
\end{equation}
where $\nu^+$ is:
\begin{equation}
    \nu^+ = \nu \frac{c + v}{c}
\end{equation}

The second term represents the radiation coming from the cloud which must be Doppler-shifted to the reference frame, the radiation cloud emits is seen by the observer at lower frequencies. However, the scattering term in the source function contains background radiation which for the calculation of $\bar{J}^{\rm{dir}}$ must be shifted as well.
\begin{equation}
    I^{\rm{source}}(\nu, \tau=0, v) = S(v) \left( 1 - \rm{e}^{-\tau_0(\nu^+)} \right)
\end{equation}
where $S(v)$ is computed using Eq. \ref{source_function_with_J_dir} but the $\bar{J}^{\rm{dir}}$ is due to the Doppler shifts computed as:
\begin{equation}
    \bar{J}^{\rm{dir}}(v) = \int^{\infty}_0 J_{\nu^-}^{\rm{inc}} \phi(\nu) d\nu \times
    (1 - {\rm e}^{-\tau_0/2})/(\tau_0/2)
\end{equation}
where $\nu^-$ is:
\begin{equation}
    \nu^- = \nu \frac{c}{c + v}
\end{equation}

The total intensity observed by the observer is a sum of these two terms:
\begin{equation}
    I^{\rm{cloud}}(\nu, \tau=0, v) = I^{\rm{attenuated}}(\nu, \tau=0, v) + I^{\rm{source}}(\nu, \tau=0, v)
\end{equation}
Using the formula \ref{arcade_intensity} we can calculate the intensity of the whole arcade.

For the purpose of understanding how the background and source terms affect the final synthetic spectrum we sum them over the whole arcade, like in Eq. \ref{arcade_intensity}:
\begin{equation}
    I^{\rm{background}}(\nu, \tau=0) = \frac{\sum^{n}_{k=1} I^{\rm{attenuated}} (\nu, \tau=0, v_{k}) A_{k}}{A^{\rm{flare}}}
\end{equation}
and:
\begin{equation}
    I^{\rm{source}}(\nu, \tau=0) = \frac{\sum^{n}_{k=1} I^{\rm{source}} (\nu, \tau=0, v_{k}) A_{k}}{A^{\rm{flare}}}
\end{equation}

\subsection{Parameter summary}
The model described above has a few parameters and they are summarized below.

We need to know the basic properties of the star on which the loop arcade forms. Parameters that describe the star in our model are stellar radius $R$ and the quiescent specific intensity $I_{\nu}$.

The macroscopic description of the loop arcade requires the radius of the semicircular loops $r$, the area of the arcade $A^{\rm{flare}}$ and the initial velocity of clouds of plasma at the top of the arcade $v_0$.

For the description of the plasma in clouds, we use the following parameters: plasma temperature $T$, the optical thickness of the clouds at the line center $\tau_0$ and turbulent velocity $v_t$ in plasma which affects the profile width $\phi(\nu)$.

\section{Results}
Results of the modeling described above are presented in this section as the spectra computed for various values of the input parameters. For the quiescent spectrum illuminating the clouds, we used a Gaussian profile which we obtained by fitting the calibrated AD Leo observation in the H$\alpha$ region plus the continuum fit. In reality, AD Leo H$\alpha$ does not have a strictly Gaussian shape but exhibits a reversal in the line core, shown in the Fig. \ref{fig:model_input_spectrum}. However, to demonstrate the effect of flows on our models we assumed that a schematic input spectrum will provide basic results - note that the wing asymmetry is mainly sensitive to the incident radiation outside the central reversal. The fitted Gaussian spectrum is shown in Fig. \ref{fig:model_input_spectrum}. The stellar radius $R$ for AD Leo we used  is 0.39~R$_{\odot}$ \citep{Reiners2009}.

\begin{figure}
    \centering
    \resizebox{\hsize}{!}{\includegraphics{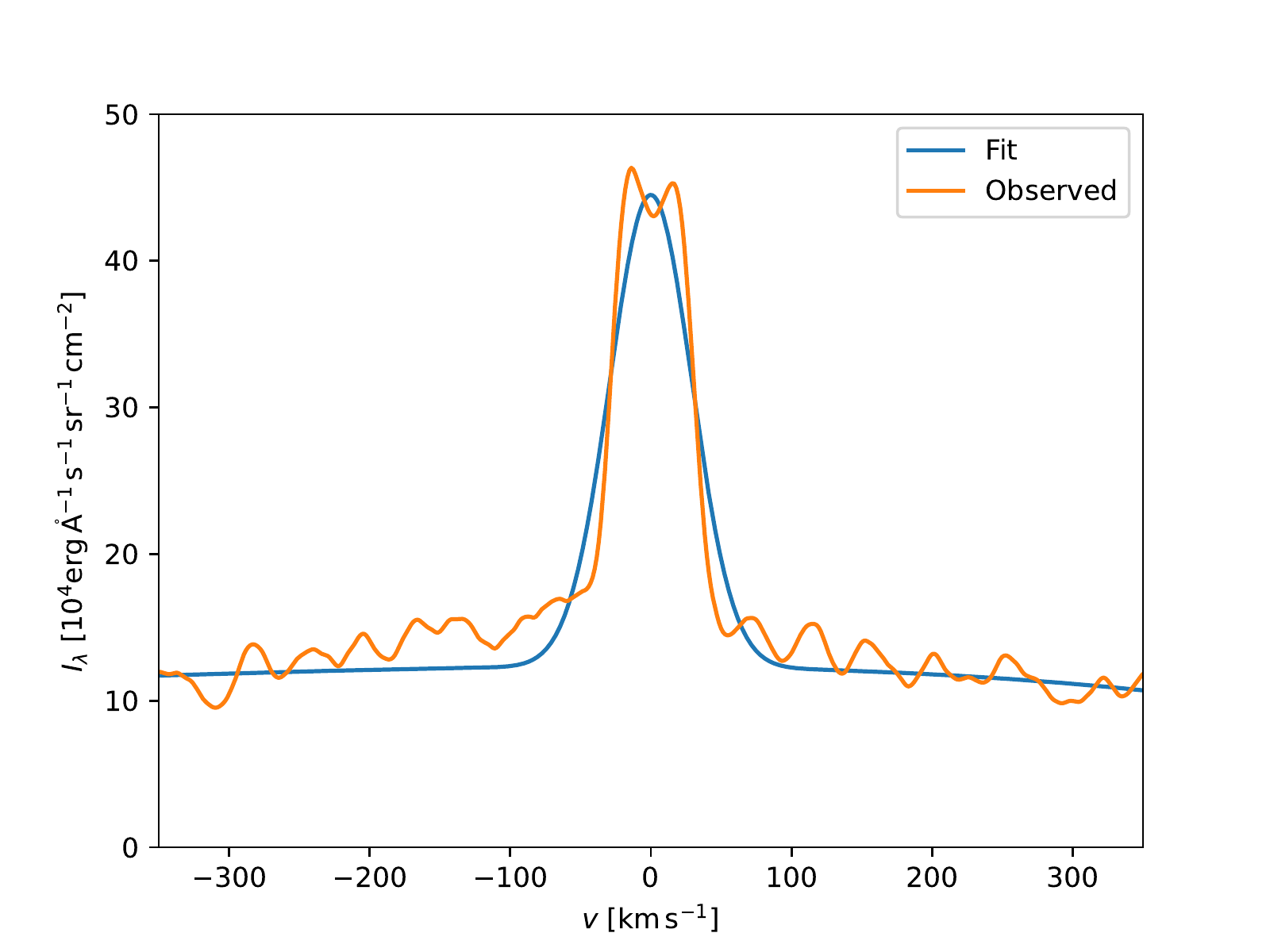}}
    \caption{The observed calibrated quiescent spectrum of the AD Leo and the fitted Gaussian profile in the H$\alpha$ region. This spectrum is used as the background radiation for our model. Compared to the observed AD Leo H$\alpha$ line this profile does not feature a small reversal in the line center.}
    \label{fig:model_input_spectrum}
\end{figure}

\subsection{Modeled profiles}

We modeled H$\alpha$ line profiles for various input parameters and the results are presented in the following grid of synthetic profiles. One parameter is varied and the rest are fixed. We study the effect of parameters that describe the plasma in our clouds. In the figures below we plot synthetic specific intensities calculated using Eq. \ref{arcade_intensity} with total source term and total attenuated background term as described above.

Temperature variations for selected parameter values are plotted in Fig. \ref{fig:model_temperature_variations}.
For lower electron density and optical thickness (a -- d panels), one can see that the source term radiation does not change much with increasing temperature. That is expected as at lower electron densities the photon destruction probability $\epsilon$ is very low so the scattering term dominates the source function.
For higher electron density (e -- h panels) the higher photon destruction probability significantly amplifies the contribution of the thermal term in the source function. Higher optical thickness attenuates the background radiation and also amplifies the source term due to diffusion.

\begin{figure*}
\centering
\begin{tabular}{cc}
  \includegraphics[width=0.38\textwidth]{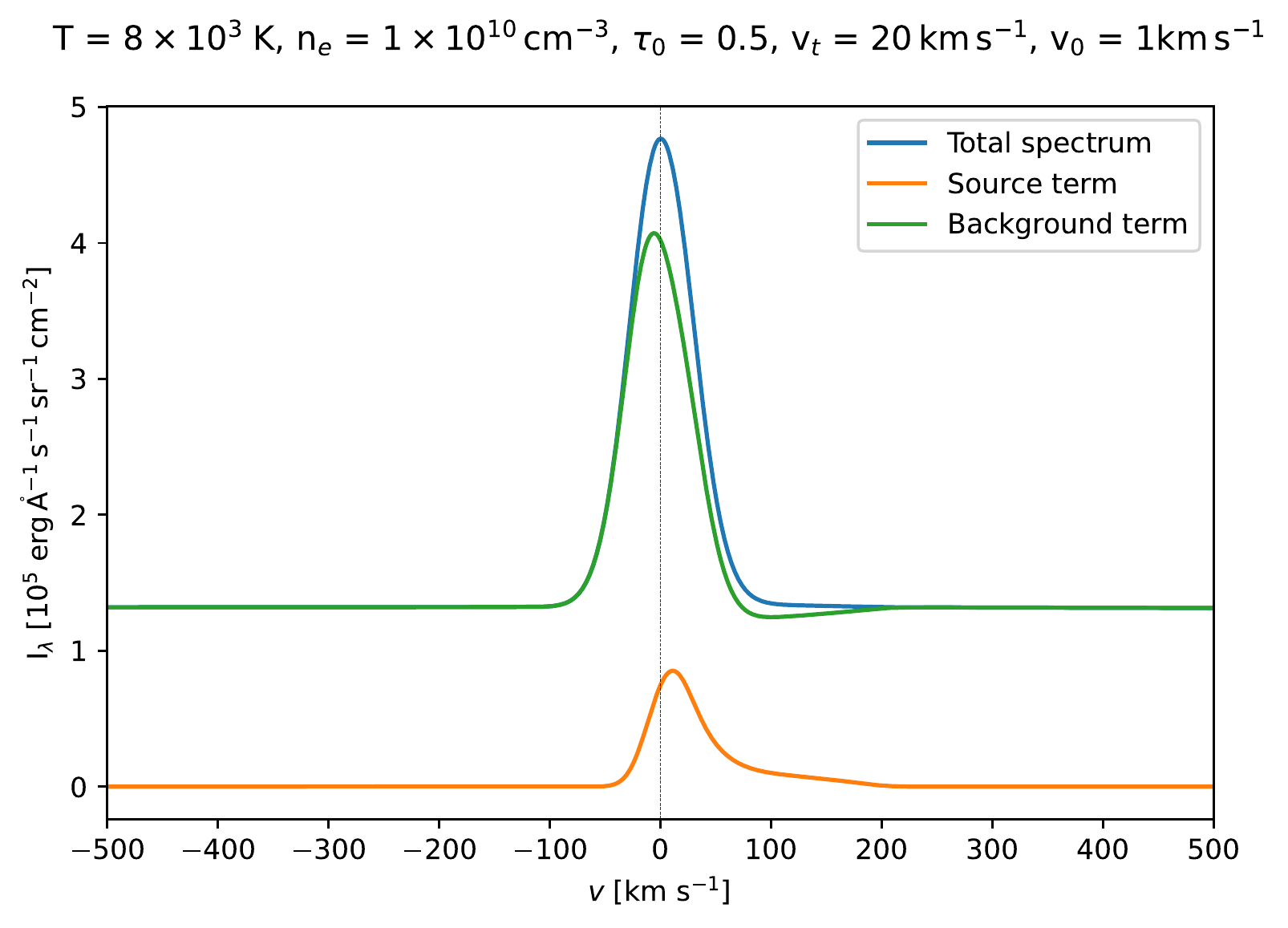} &   \includegraphics[width=0.38\textwidth]{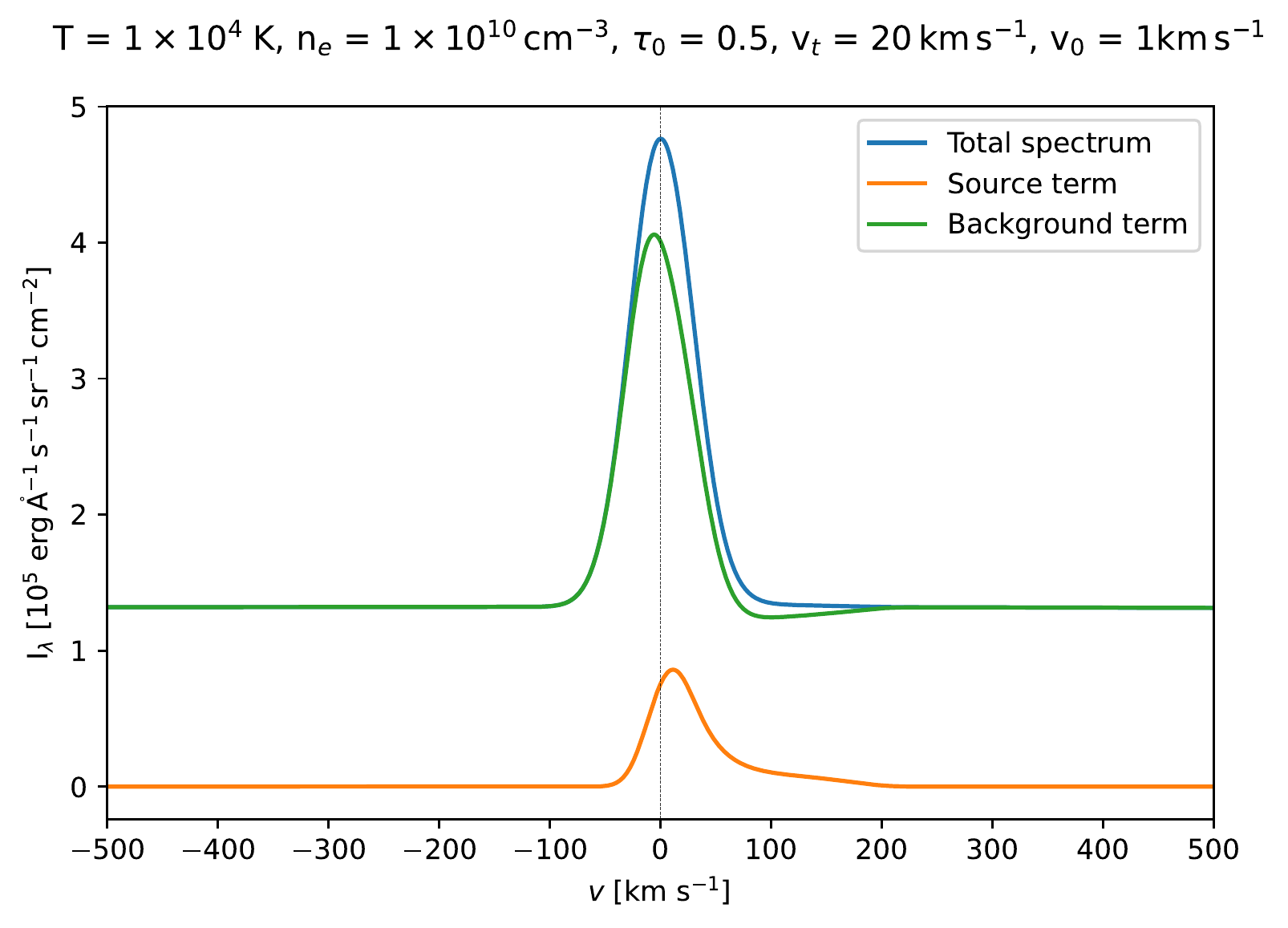} \\
(a) & (b) \\[6pt]
 \includegraphics[width=0.38\textwidth]{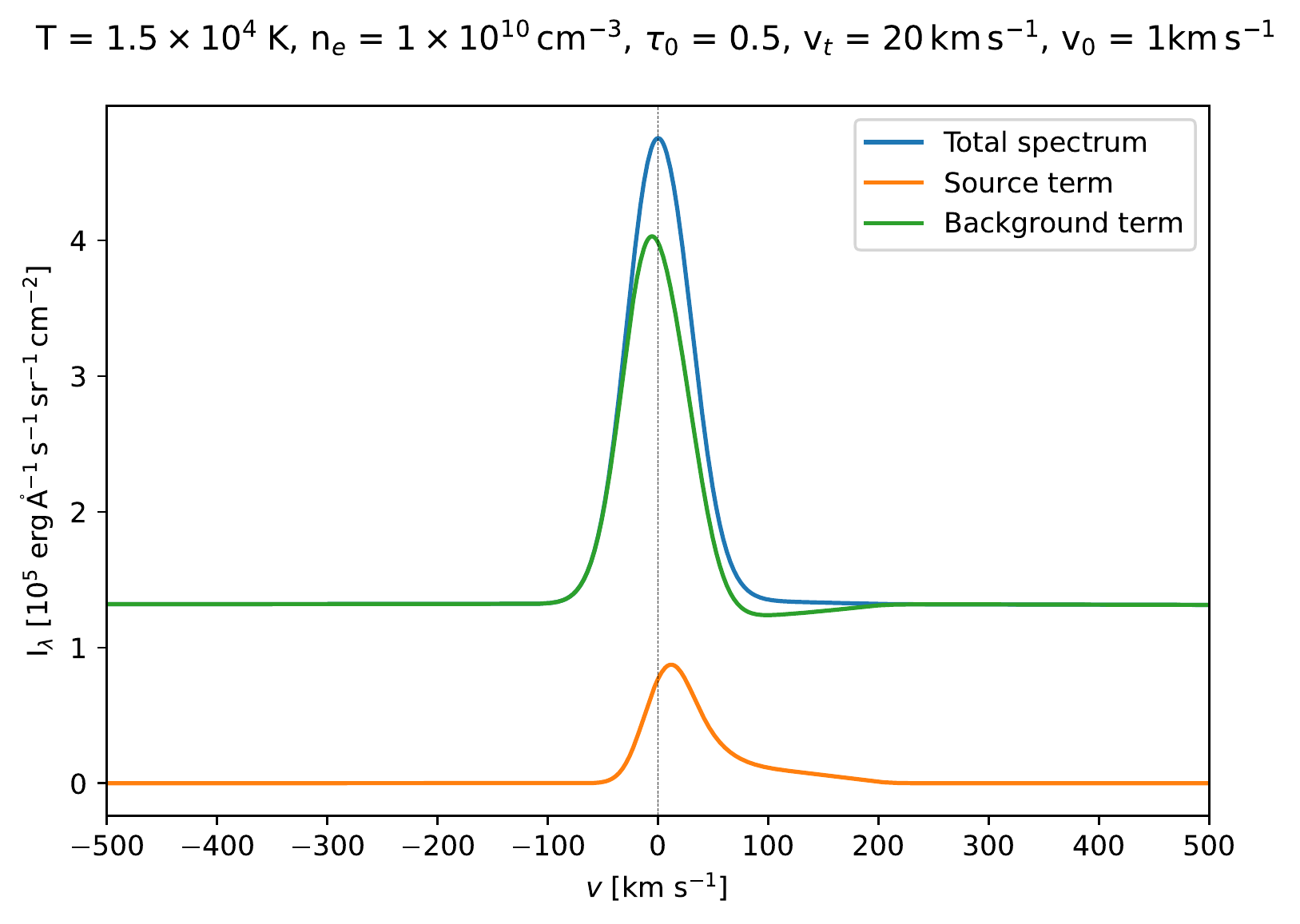} &   \includegraphics[width=0.38\textwidth]{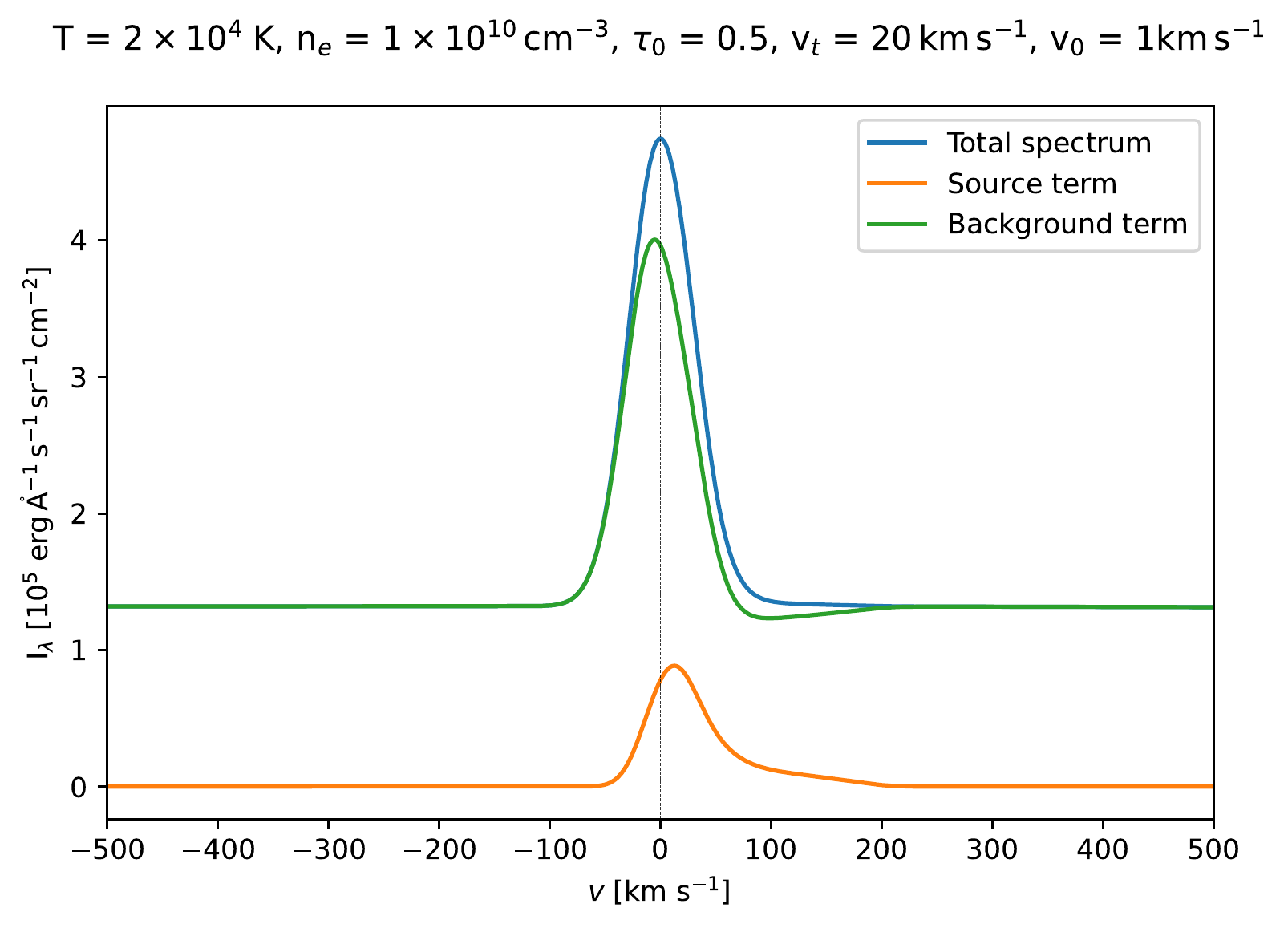} \\
(c) & (d) \\[6pt]
\includegraphics[width=0.38\textwidth]{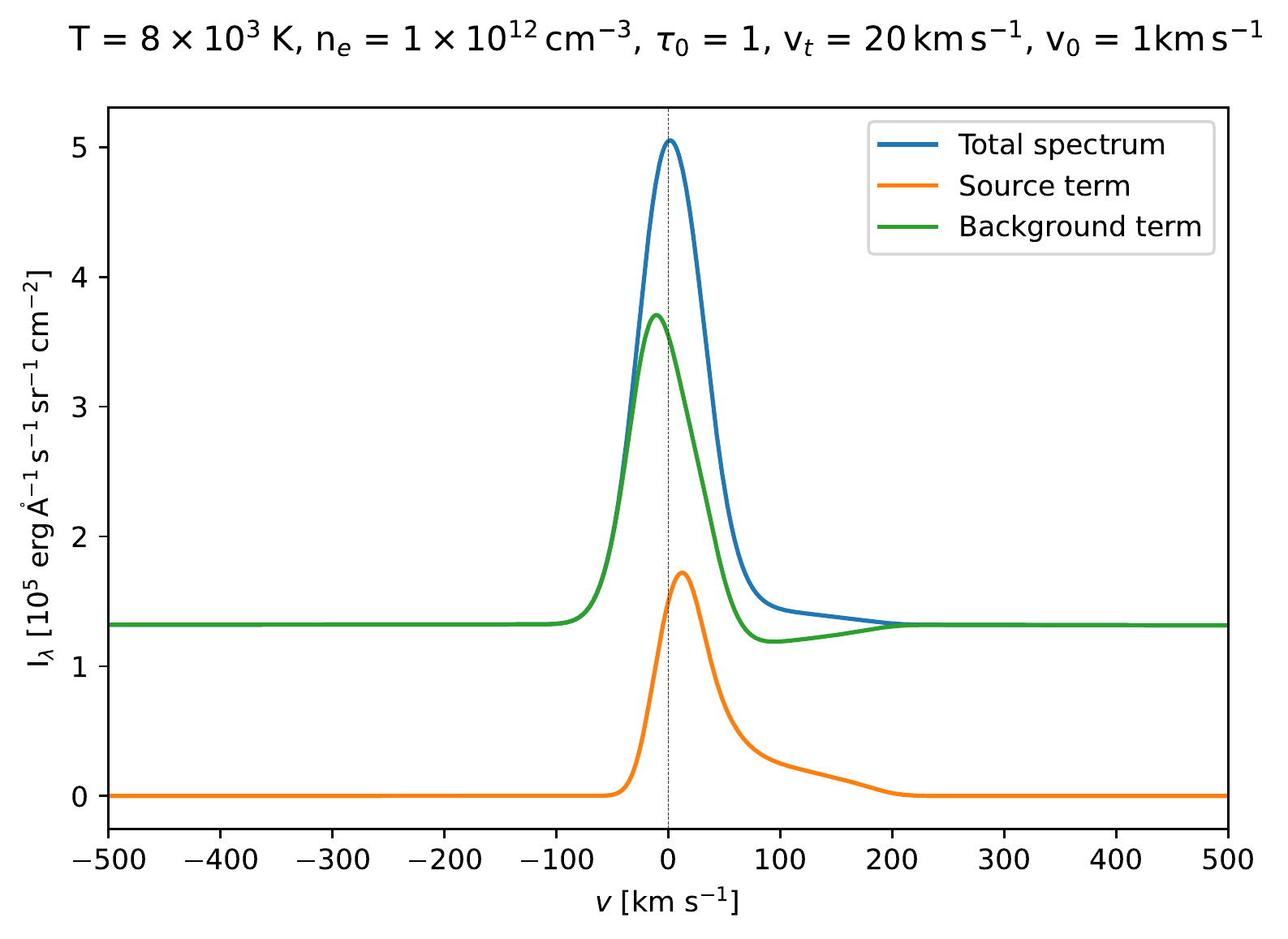} &   \includegraphics[width=0.38\textwidth]{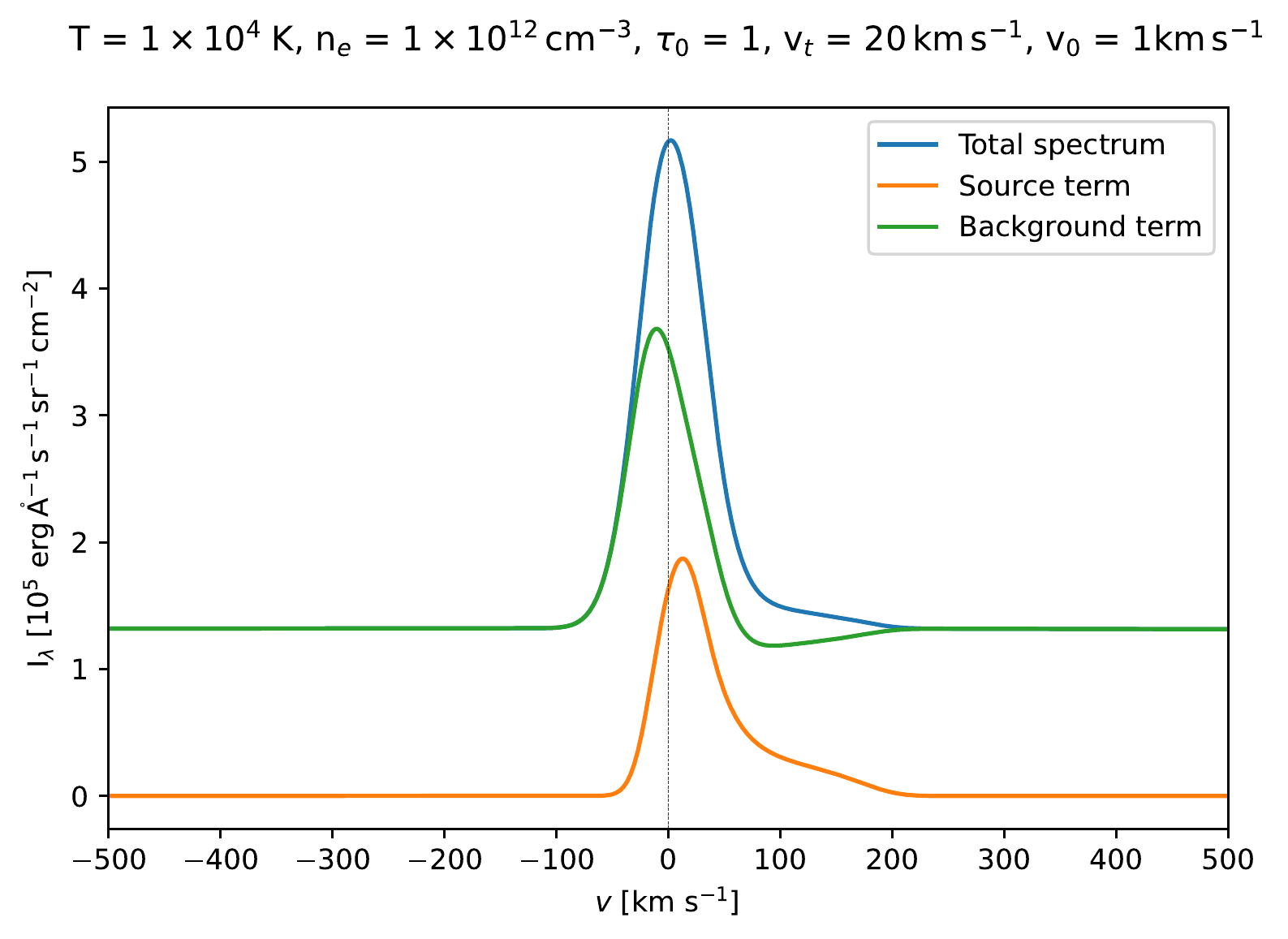} \\
(e) & (f) \\[6pt]
\includegraphics[width=0.38\textwidth]{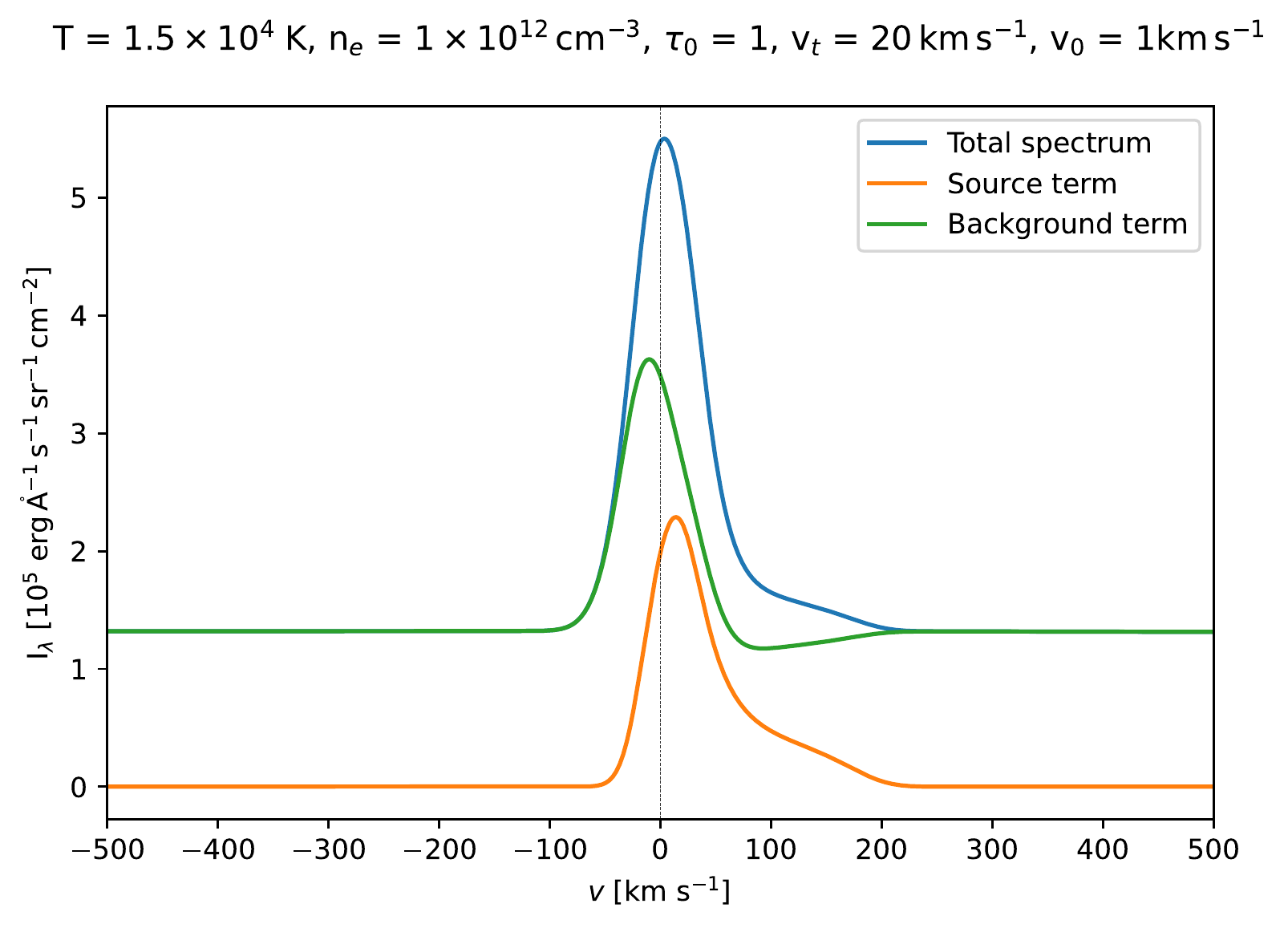} &   \includegraphics[width=0.38\textwidth]{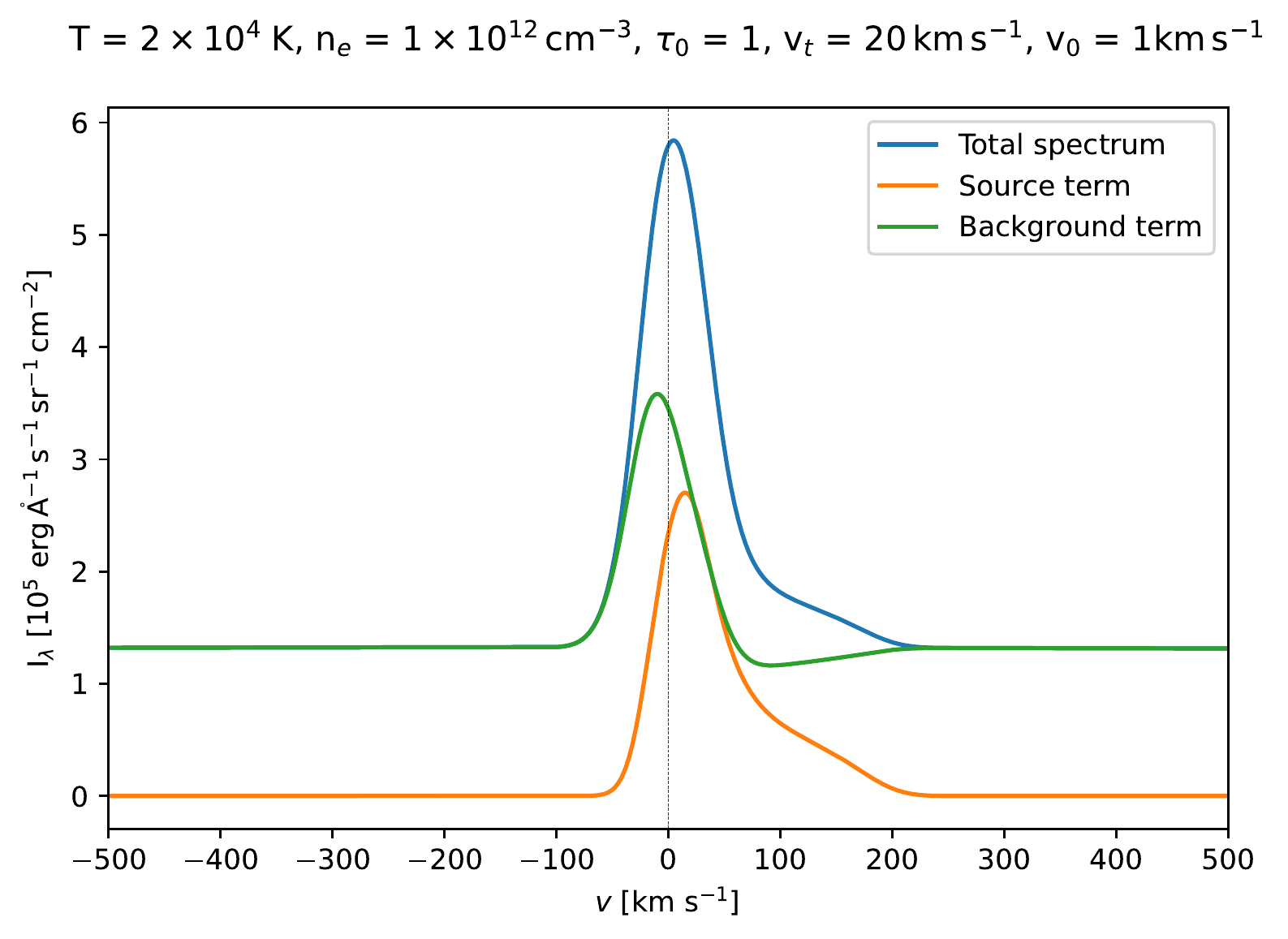} \\
(g) & (h) \\[6pt]
\end{tabular}
\caption{Synthetic H$\alpha$ line profiles. Modeled specific intensity is a sum of attenuated background radiation and source radiation. Both components are plotted as well as the sum. This figure shows variations of the modeled profiles with temperature for two values of electron density and optical thickness.}
\label{fig:model_temperature_variations}
\end{figure*}

Electron density variations for selected parameter values are plotted in Fig. \ref{fig:model_electron_density_variations}.
For lower temperature and higher optical thickness panels (a -- d) show that the source term is significantly more amplified with increasing electron density.
For higher temperature and lower optical thickness (e -- h panels), we can see that the source term radiation is increasing again with increasing electron density. We use the electron densities up to the relatively high $10^{13}$ $\rm{cm^{-3}}$. These electron densities have been recently found in cool flare loops on Sun, see \citet{Jejcic2018}.

\begin{figure*}
\centering
\begin{tabular}{cc}
  \includegraphics[width=0.38\textwidth]{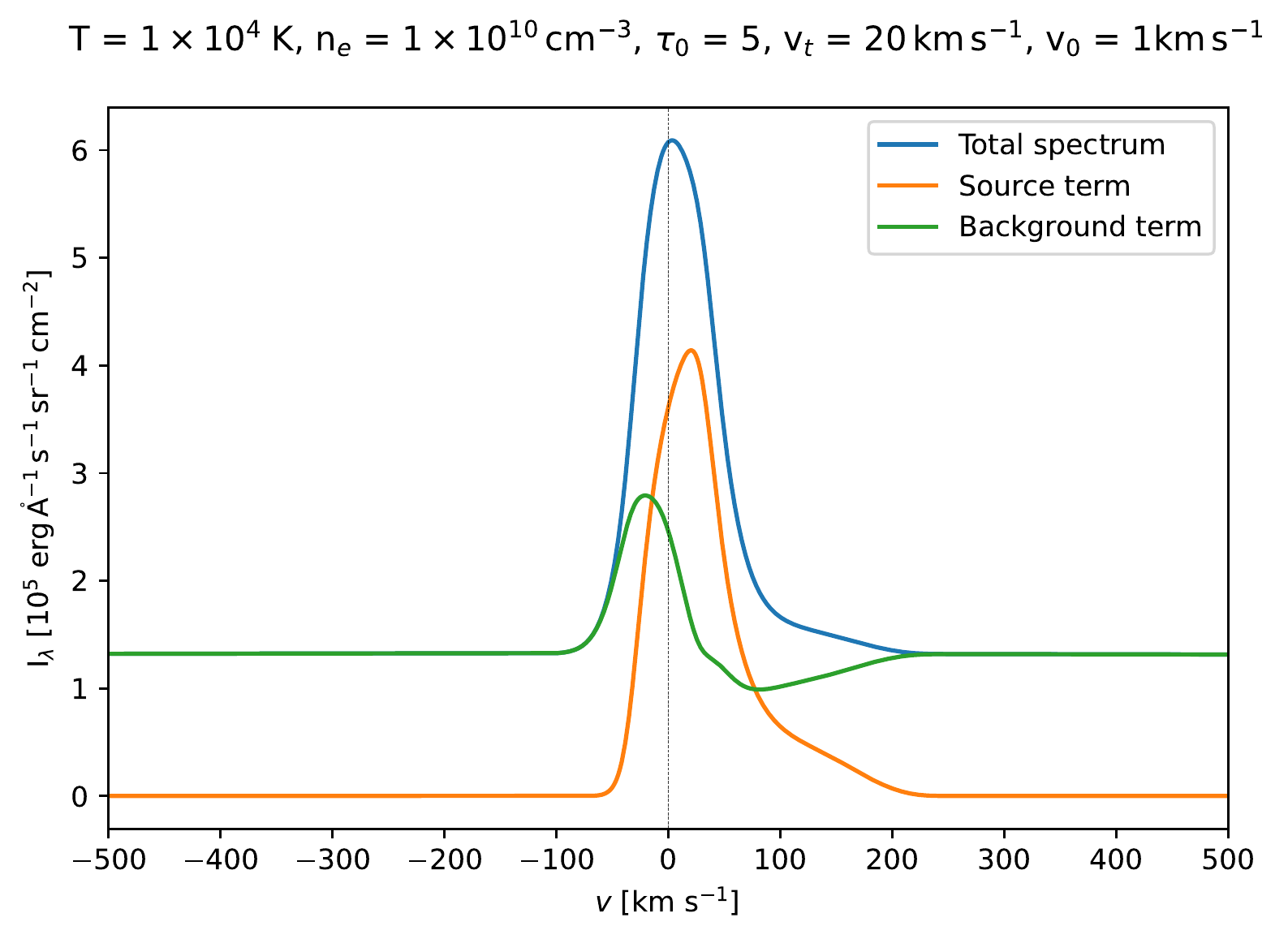} &   \includegraphics[width=0.38\textwidth]{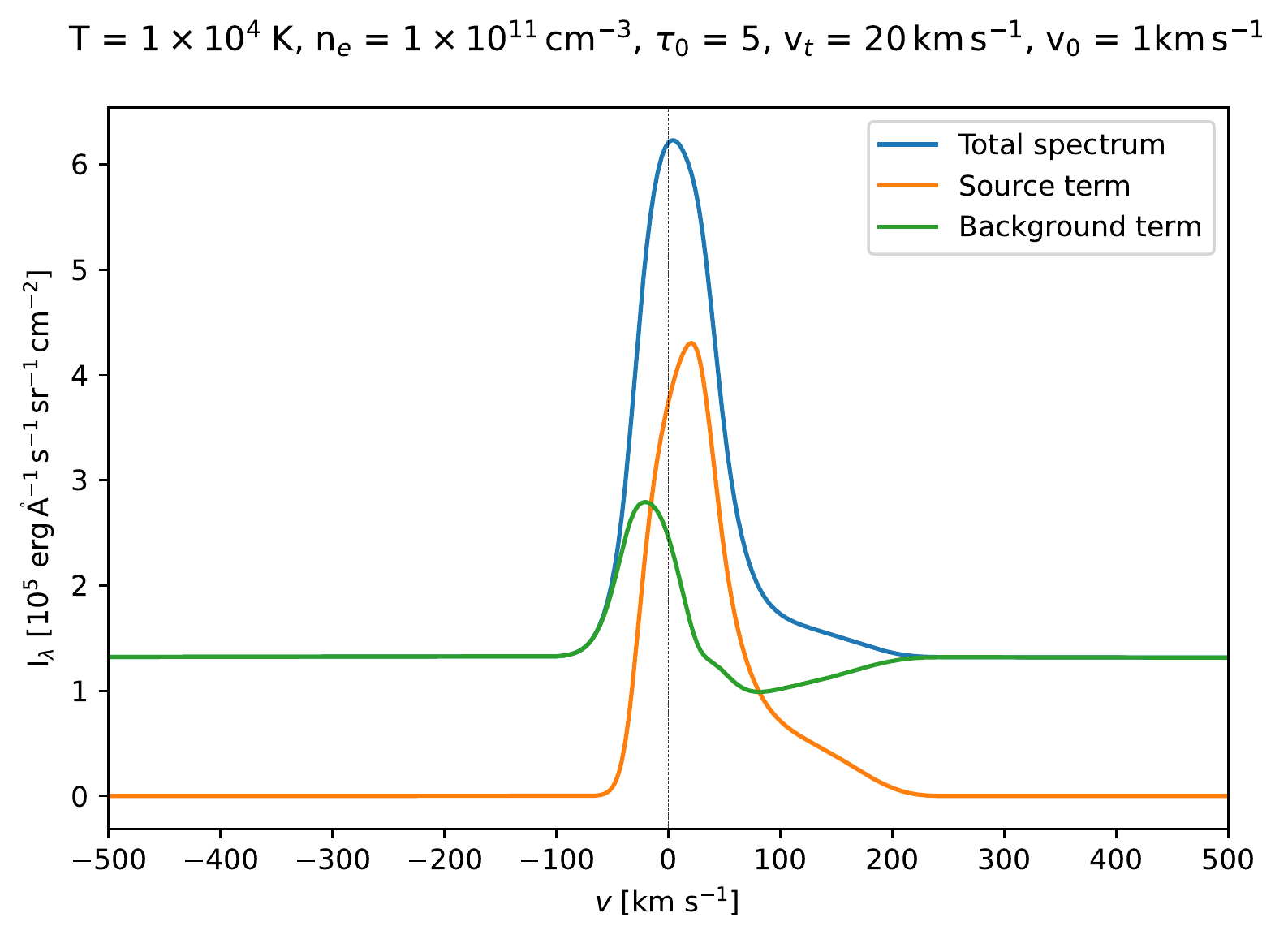} \\
(a) & (b) \\[6pt]
 \includegraphics[width=0.38\textwidth]{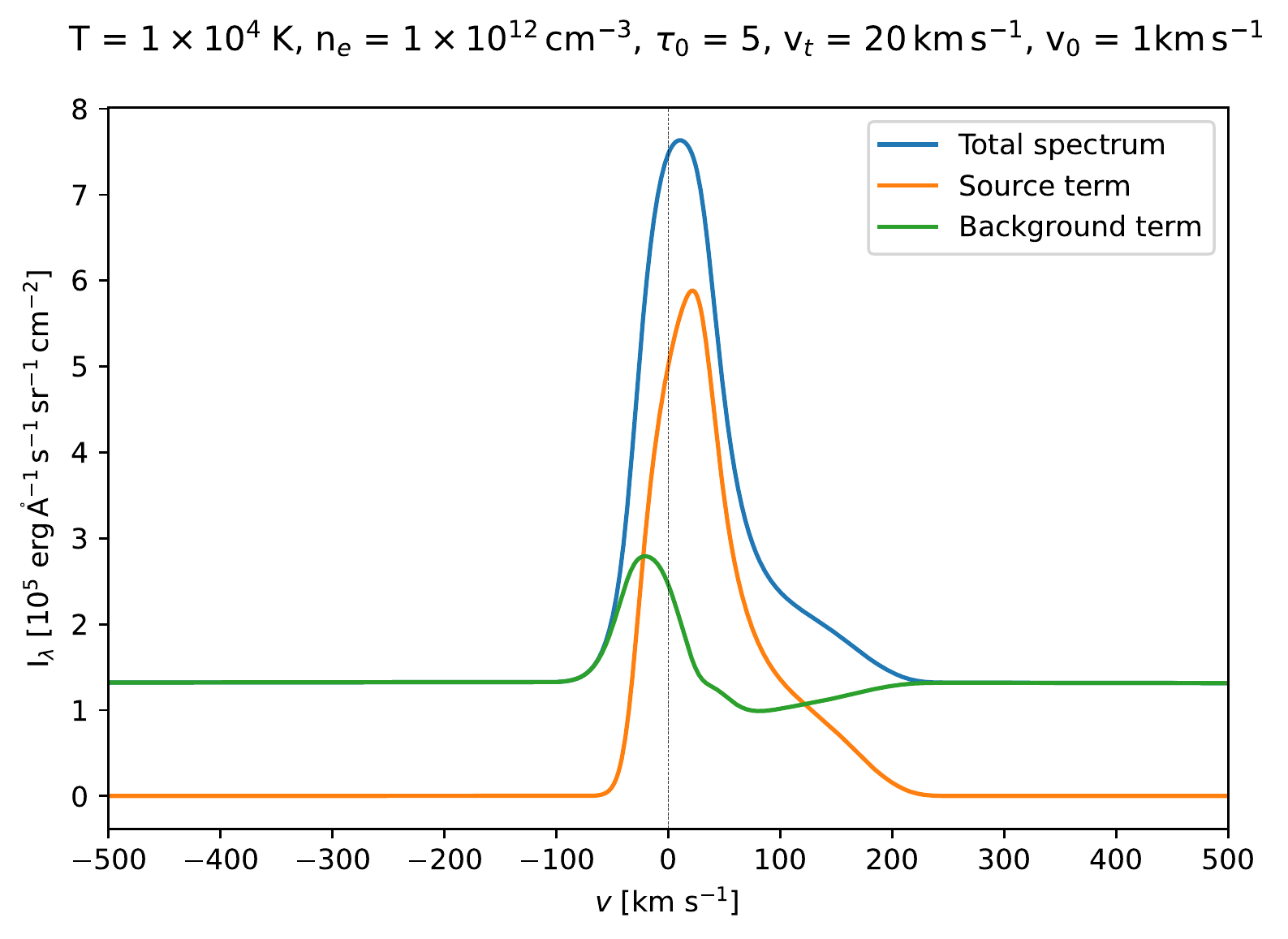} &   \includegraphics[width=0.38\textwidth]{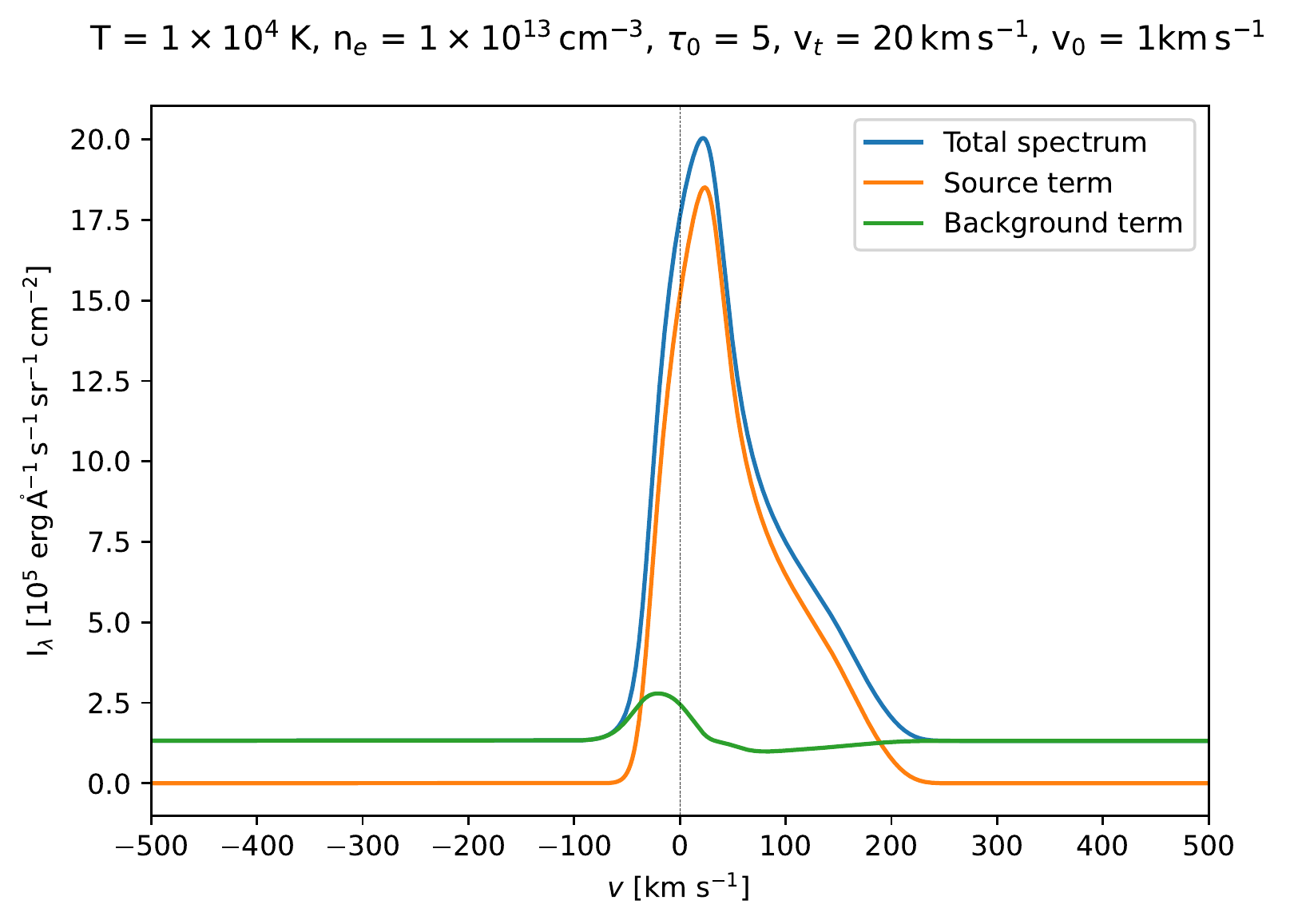} \\
(c) & (d) \\[6pt]
  \includegraphics[width=0.38\textwidth]{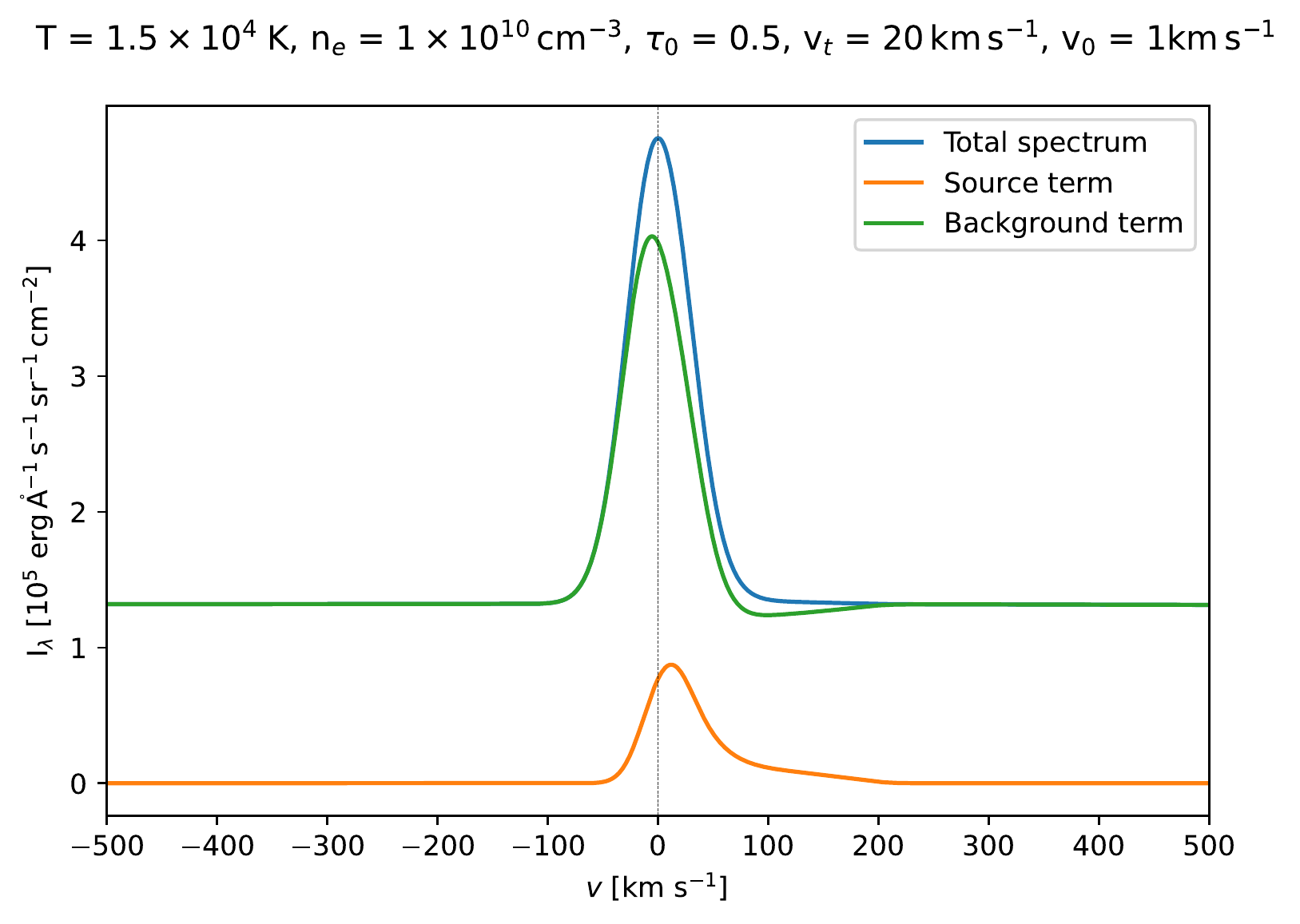} &   \includegraphics[width=0.38\textwidth]{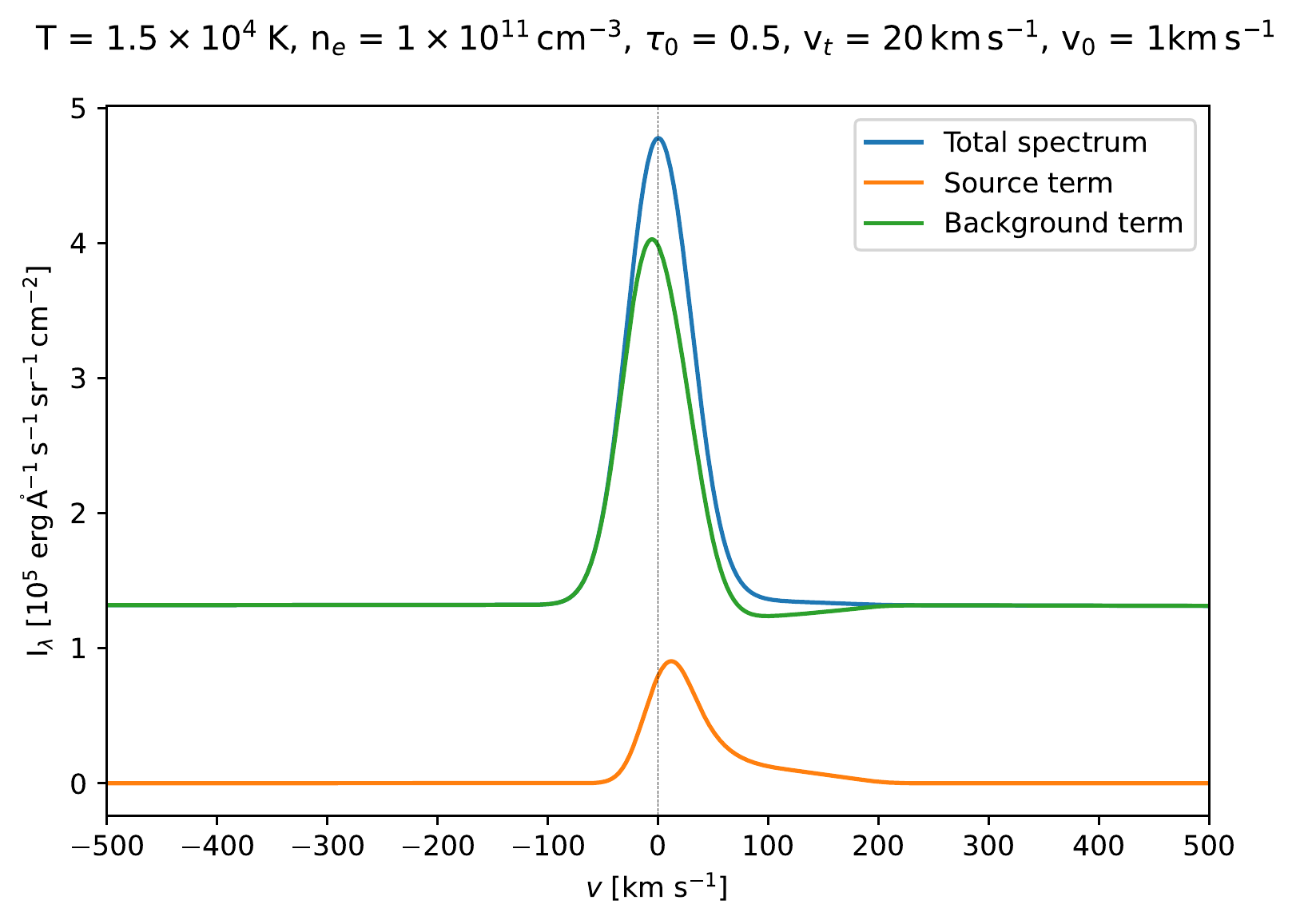} \\
(e) & (f) \\[6pt]
 \includegraphics[width=0.38\textwidth]{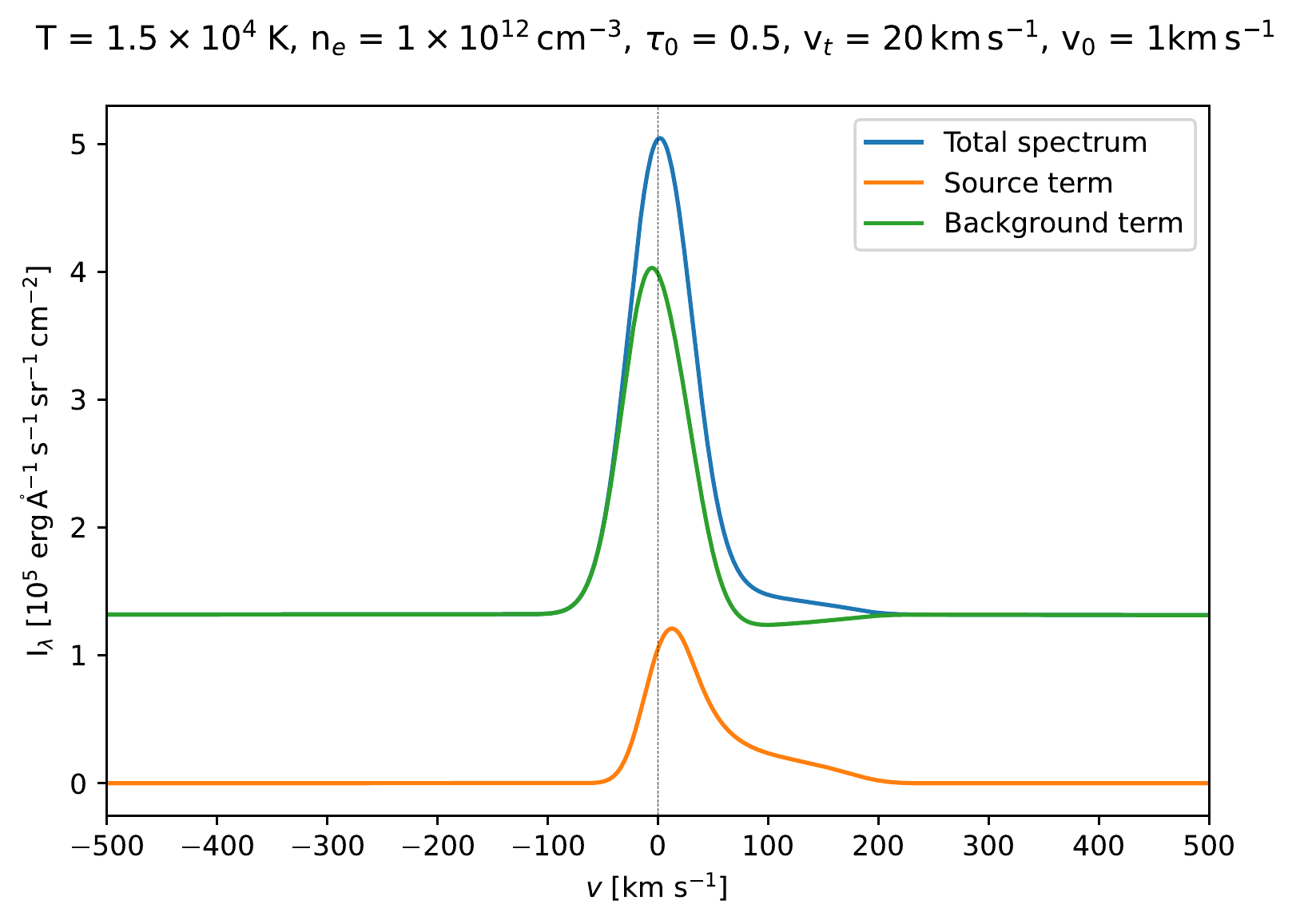} &   \includegraphics[width=0.38\textwidth]{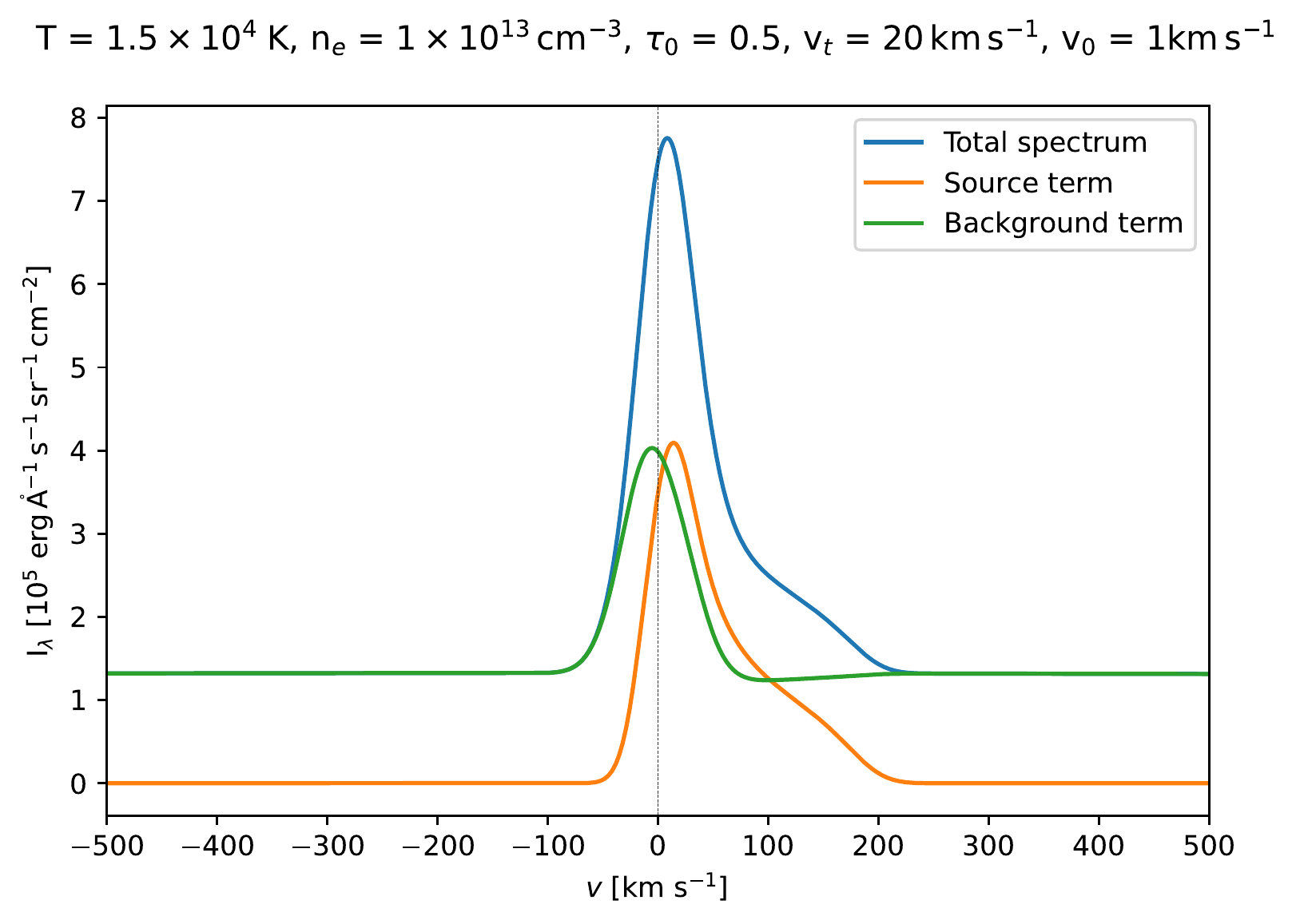} \\
(g) & (h) \\[6pt]
\end{tabular}
\caption{Synthetic H$\alpha$ line profiles. Modeled specific intensity is a sum of attenuated background radiation and source radiation. Both components are plotted as well as the sum. This figure shows variations of the modeled profiles with electron density for two values of temperature and optical thickness.}
\label{fig:model_electron_density_variations}
\end{figure*}

Optical thickness variations for selected parameter values are plotted in Fig. \ref{fig:model_optical_depth_variations}.
For low temperature (a -- c panels) we can see that the source term increases a lot and the attenuated background term decreases with increasing optical thickness.

\begin{figure*}
\centering
\begin{tabular}{cc}
  \includegraphics[width=0.4\textwidth]{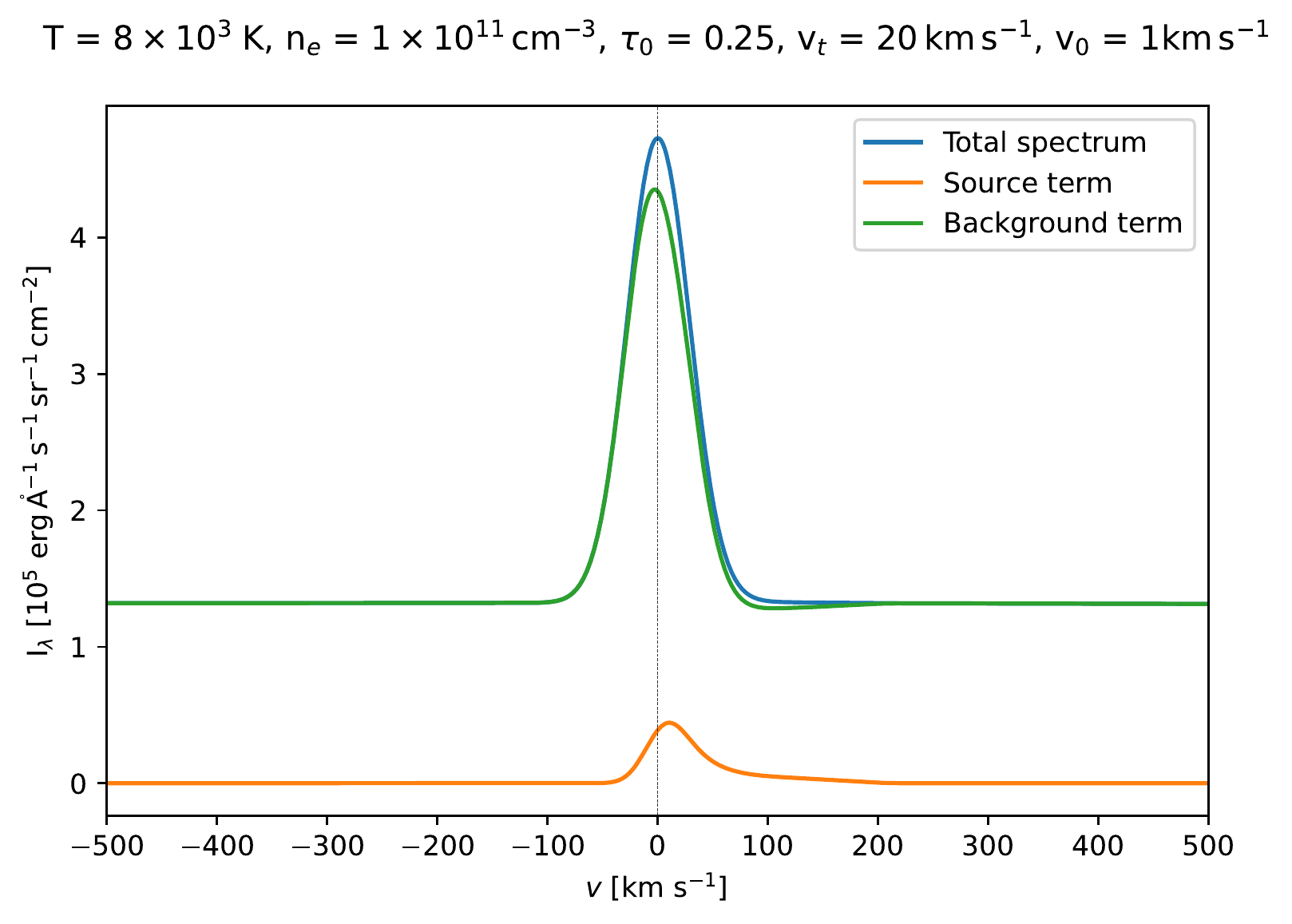} &   \includegraphics[width=0.4\textwidth]{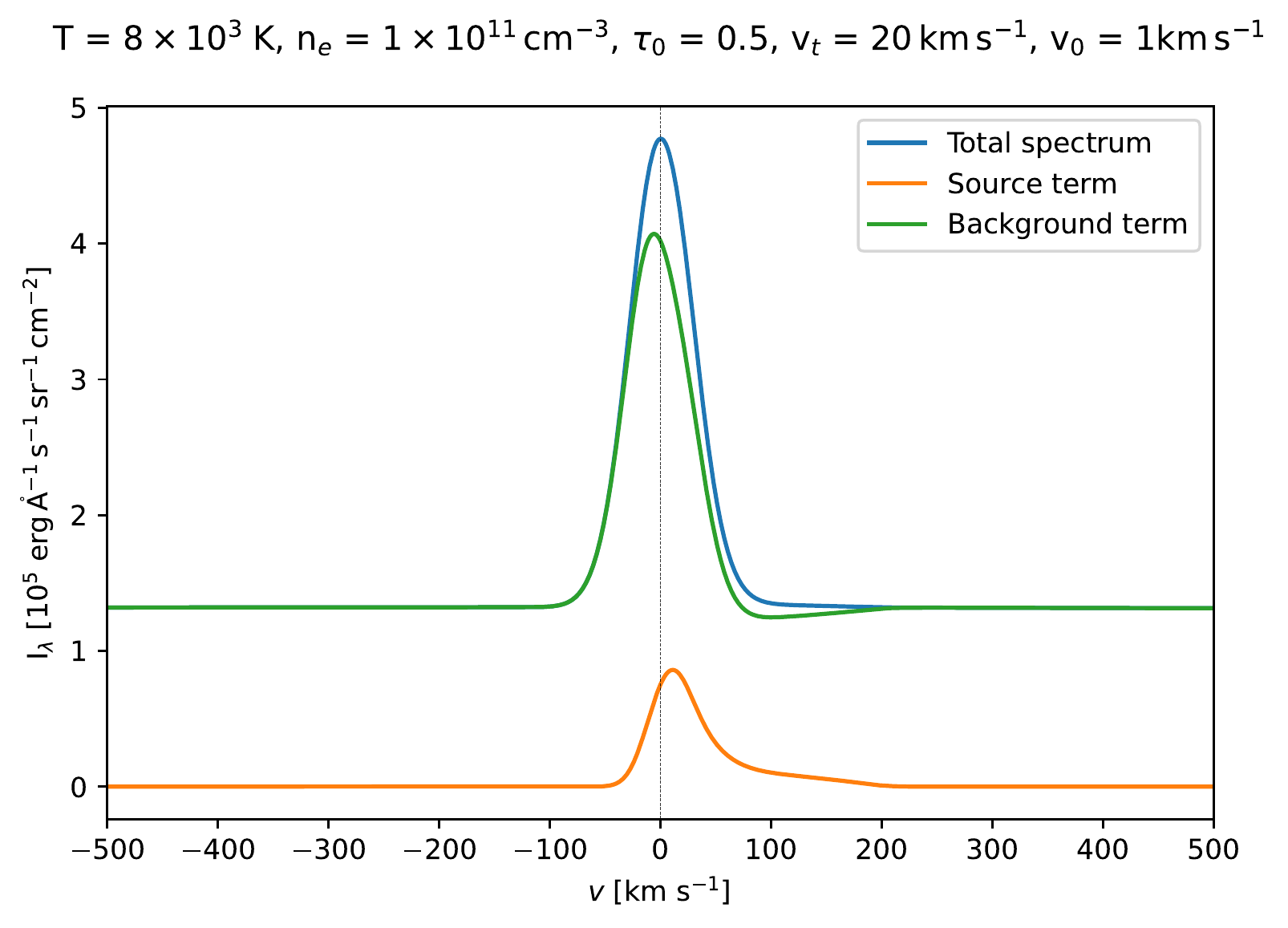} \\
(a) & (b) \\[6pt]
 \includegraphics[width=0.4\textwidth]{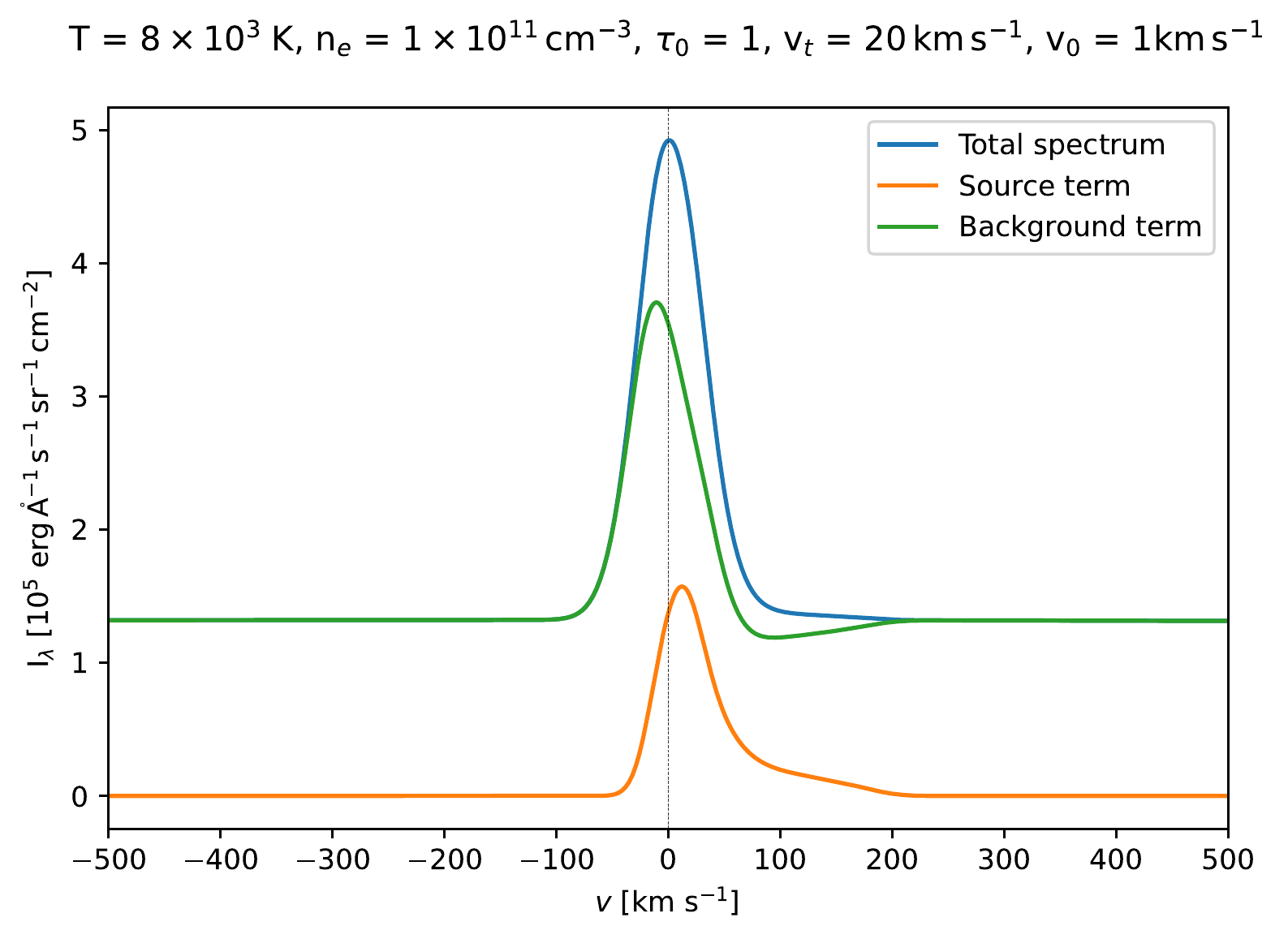} &  \includegraphics[width=0.4\textwidth]{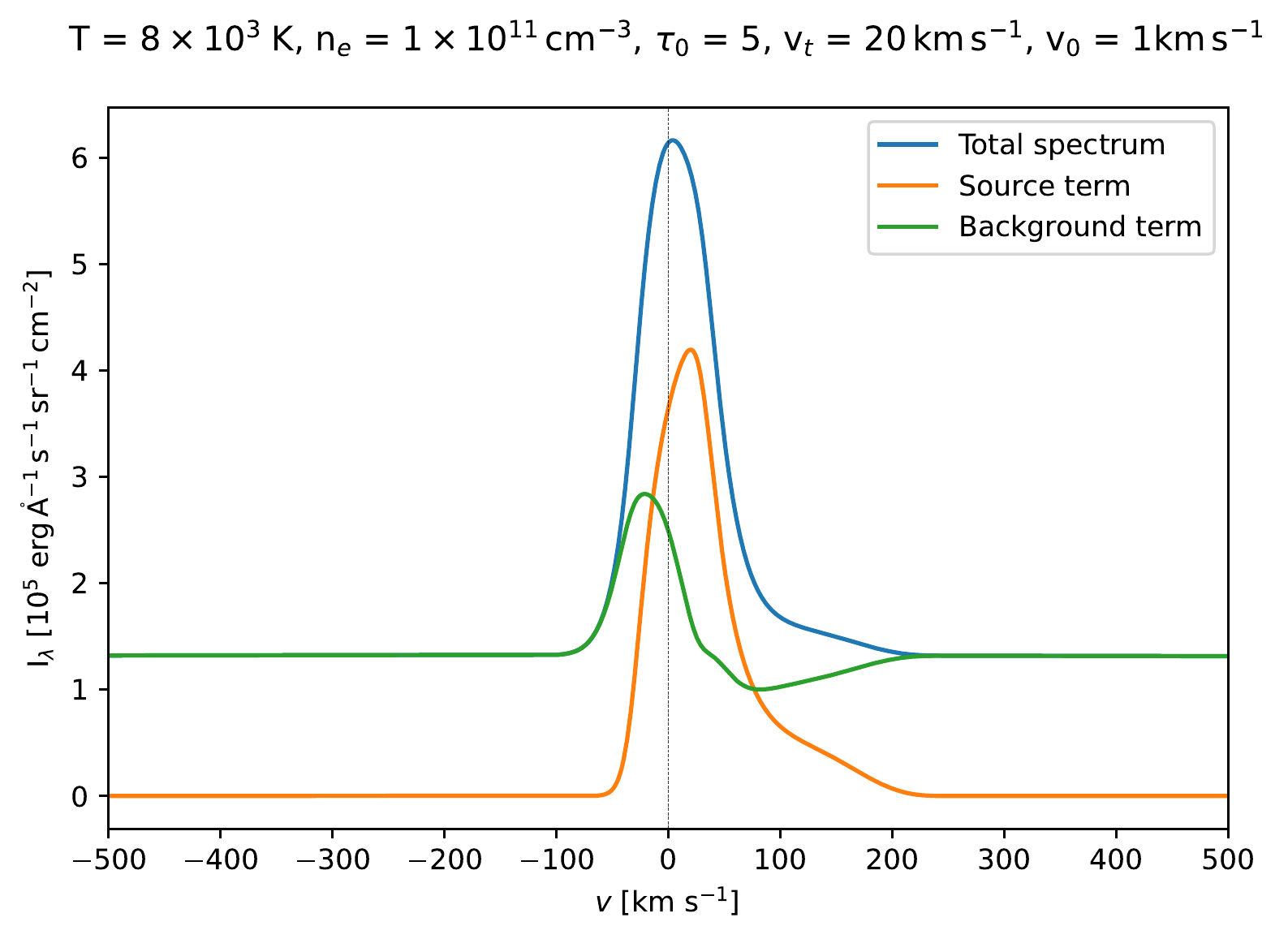}\\
(c) & (d) \\[6pt]
\includegraphics[width=0.4\textwidth]{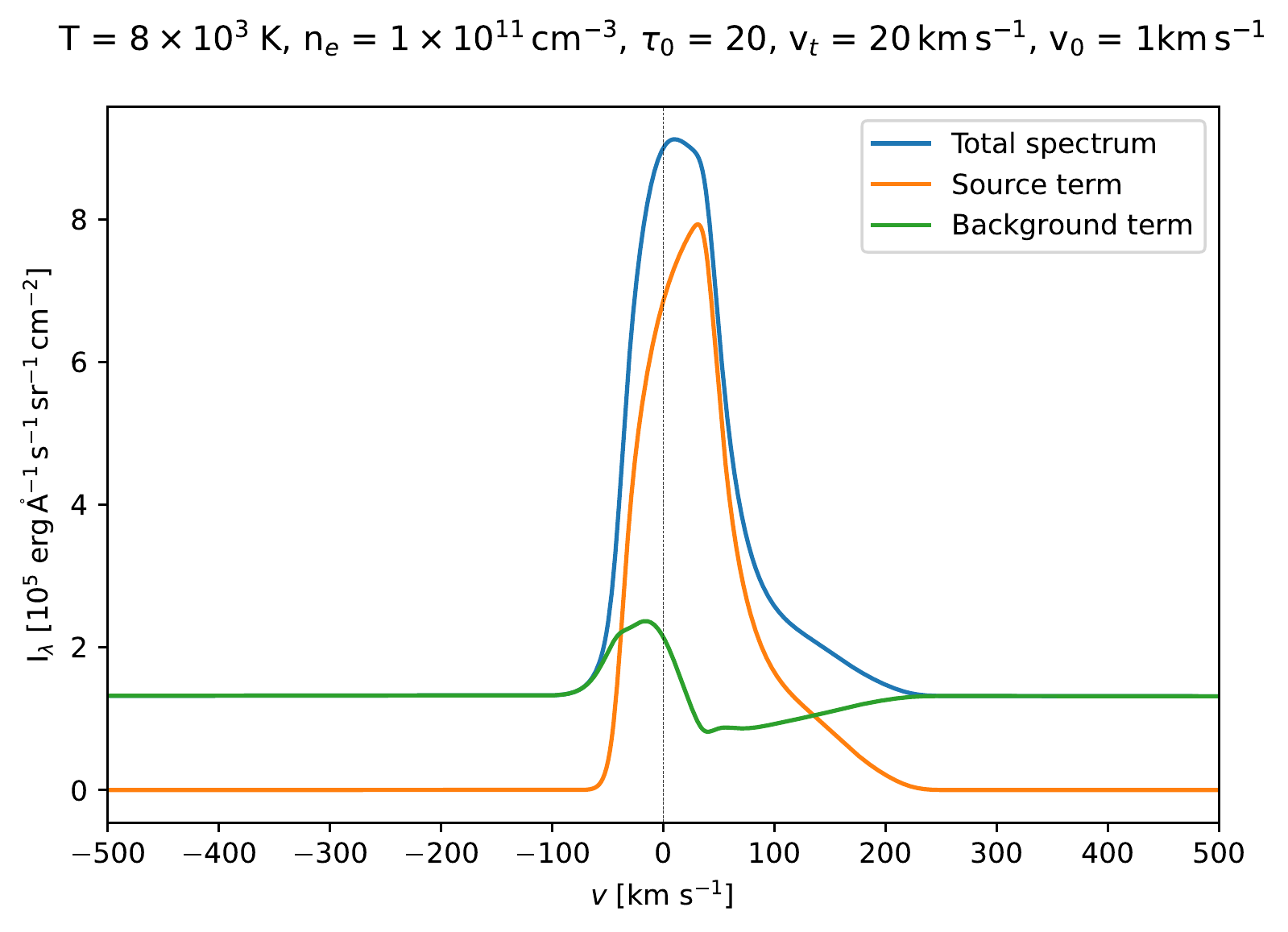} & \\
(e) & \\[6pt]
\end{tabular}
\caption{Synthetic H$\alpha$ line profiles. Modeled specific intensity is a sum of attenuated background radiation and source radiation. Both components are plotted as well as the sum. This figure shows variations of the modeled profiles with optical thickness.}
\label{fig:model_optical_depth_variations}
\end{figure*}

Turbulent velocity variations for selected parameter values are plotted in Fig. \ref{fig:model_turbulent_velocity_variations}. We used relatively large values of $v_{\rm t}$, up to 50~km\,s$^{-1}$, which is consistent with observations of \citet{Mikula2017}.
The results of our modeling indicate that increasing turbulent velocity significantly widens red wing enhancement for both lower temperature and higher optical thickness, as well as for higher temperature and lower optical thickness.

\begin{figure*}
\centering
\begin{tabular}{cc}
  \includegraphics[width=0.42\textwidth]{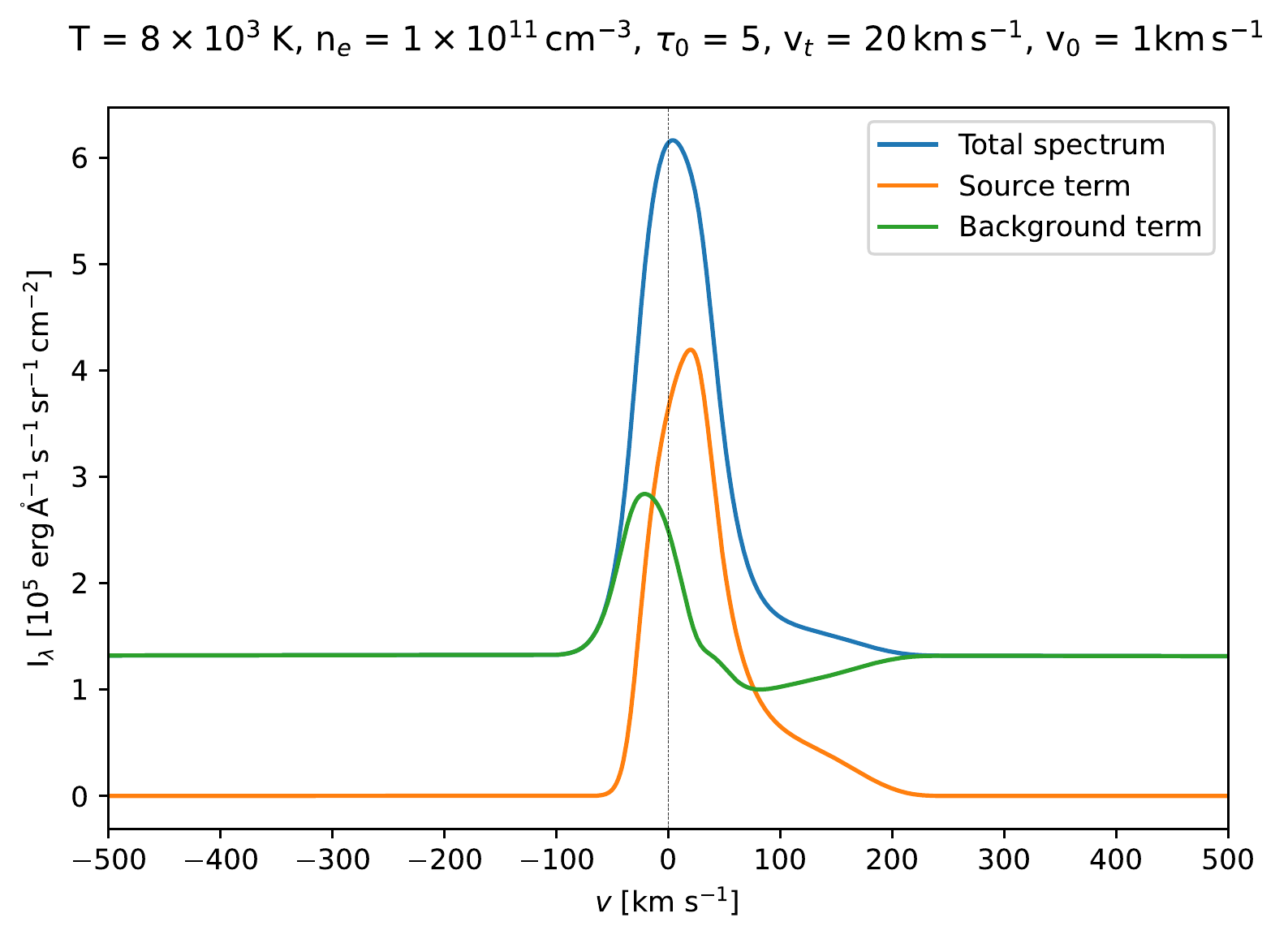} &   \includegraphics[width=0.42\textwidth]{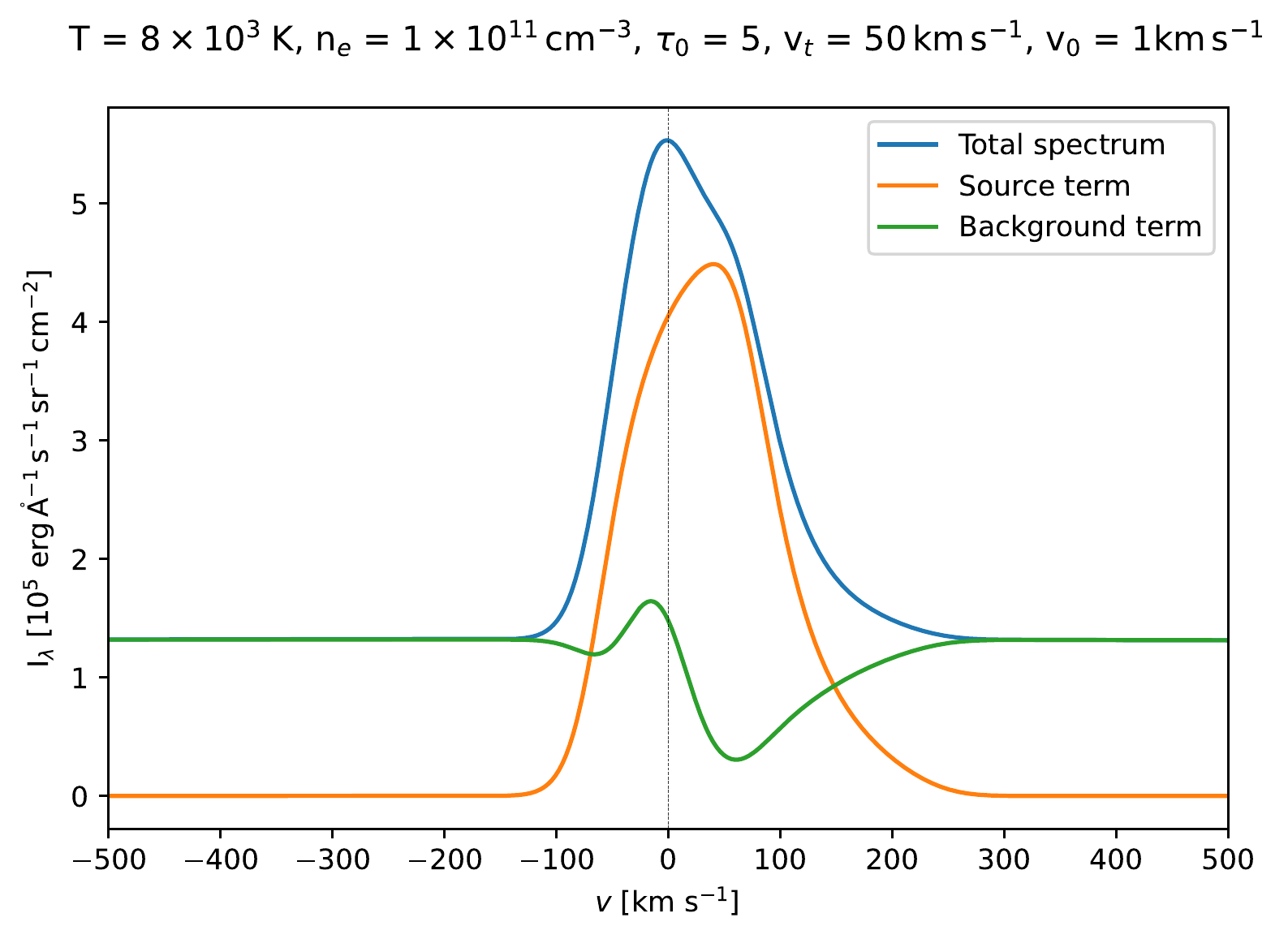} \\
(a) & (b) \\[6pt]
\includegraphics[width=0.42\textwidth]{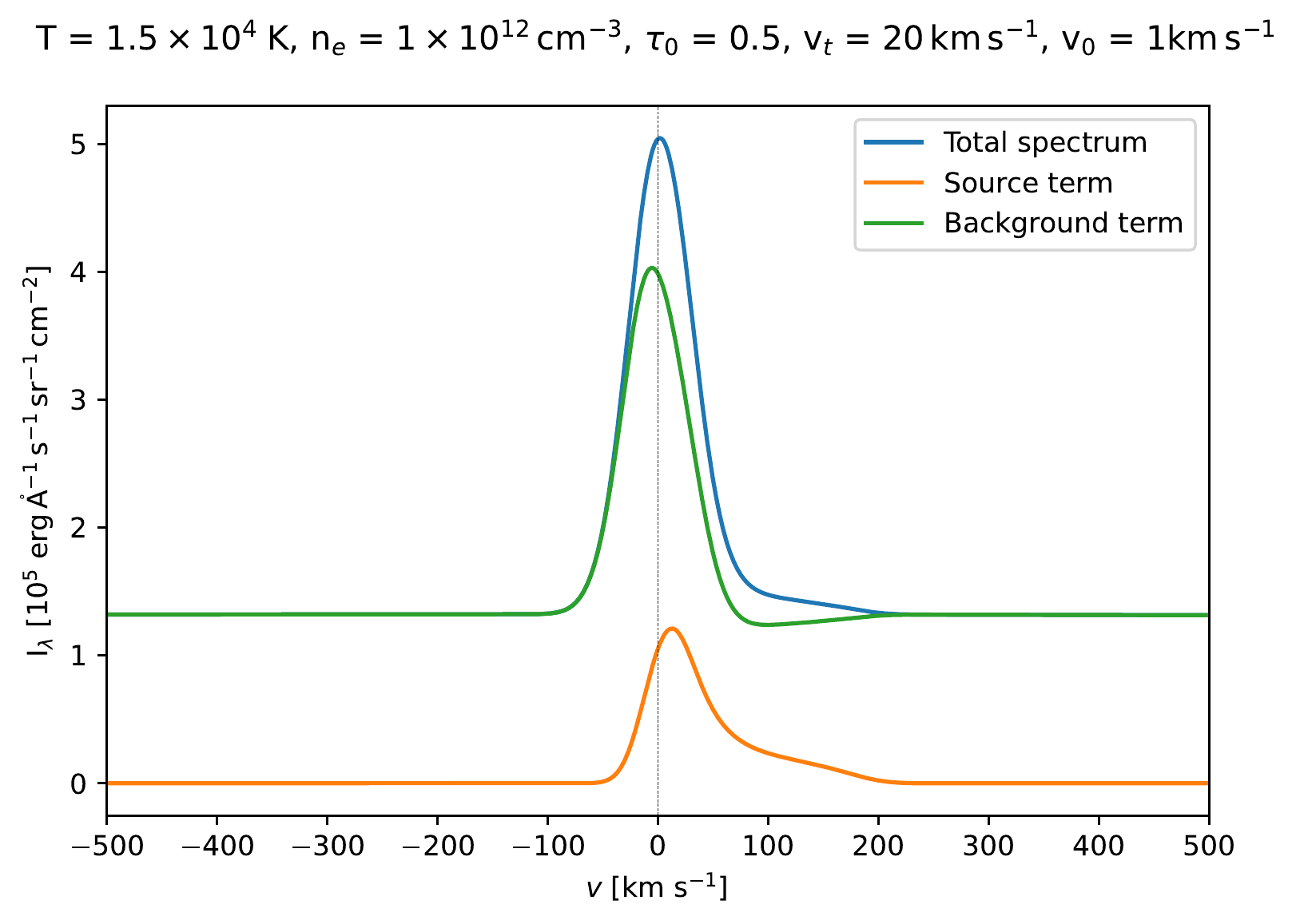} &   \includegraphics[width=0.42\textwidth]{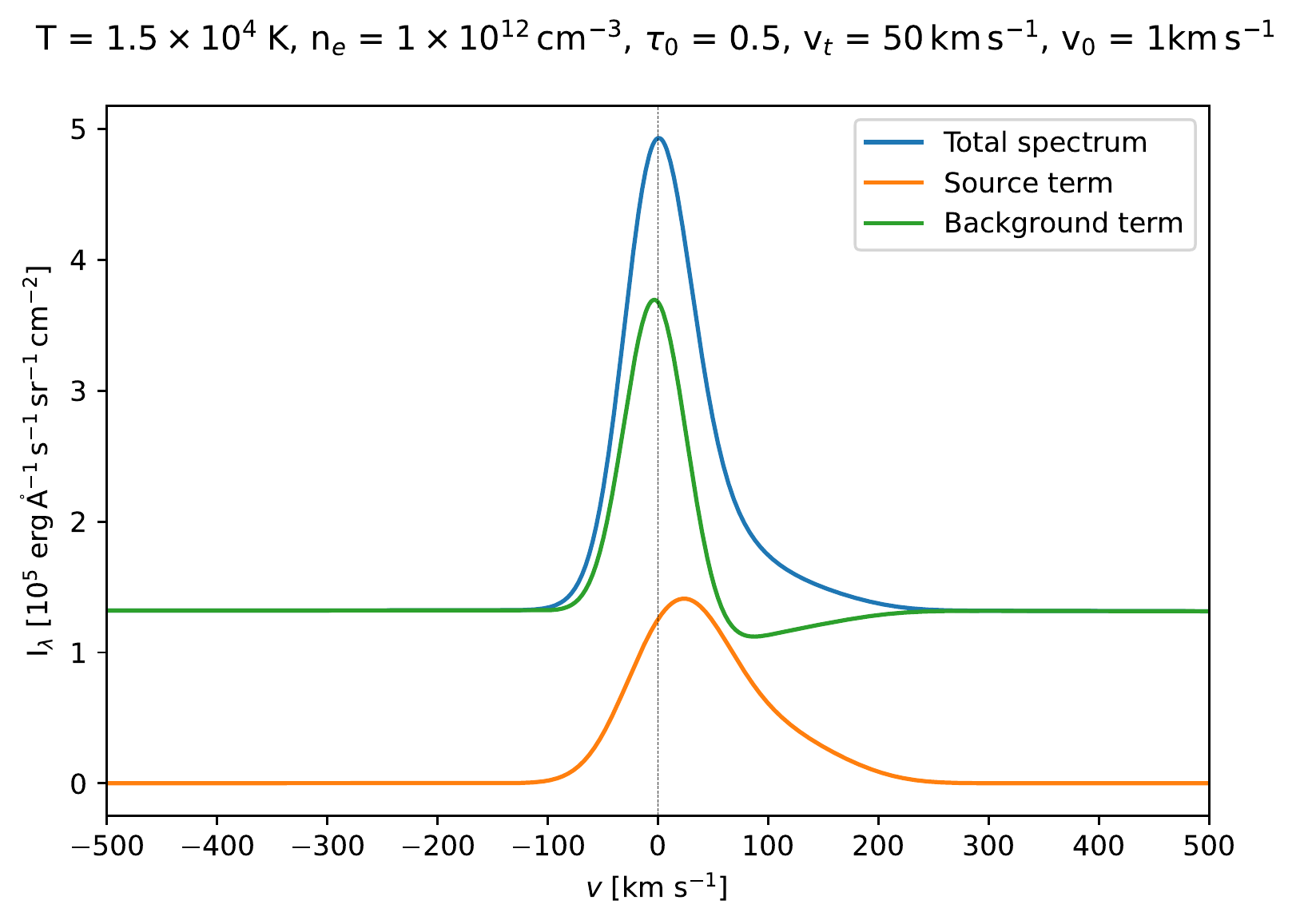} \\
(c) & (d) \\[6pt]
\end{tabular}
\caption{Synthetic H$\alpha$ line profiles. Modeled specific intensity is a sum of attenuated background radiation and source radiation. Both components are plotted as well as the sum. This figure shows variations of the modeled profiles with turbulent velocity for two values of temperature, electron density and optical thickness.}
\label{fig:model_turbulent_velocity_variations}
\end{figure*}

\subsection{Comparison with OES observations}
To test the model we attempt to fit the synthetic profile differences to our OES observations. The data presented in Fig. \ref{fig:profile_changes_2} and \ref{fig:profile_changes_1} is the difference between stellar flux during the flare and quiescent flux of AD Leo. To calculate the same effect using our model we compute flux seen by the observer on Earth and we subtract the quiescent radiation we used as the input for our model expressed as flux seen on Earth. Flux seen by an observer on Earth under all the assumptions of our model can be approximated as:
\begin{equation}
    F_{\lambda} = \frac{I^{\rm{flare}}_{\lambda} A^{\rm{flare}} + I^{\rm{quiescent}}_{\lambda} \left ( A^{\rm{disc}} - A^{\rm{flare}} \right )}{d^2}
\end{equation}
where $I^{\rm flare}_{\lambda}$ is the synthetic specific intensity calculated using Eq. \ref{arcade_intensity}, $I^{\rm quiescent}_{\lambda}$ is the quiescent specific intensity of the star, $A^{\rm flare}$ is the area of the flare, $A^{\rm disc}$ is the area of the stellar disc and $d$ is the distance from the star to the Earth.
Subtracting the quiescent flux seen by the observer we get the formula
\begin{equation}
    \Delta F_{\lambda} = \frac{\left ( I_{\lambda}^{\rm{flare}} - I_{\lambda}^{\rm{quiescent}} \right ) A^{\rm{flare}}}{d^2}
\end{equation}
simulating the formula in Eq. \ref{flux_difference}.
During fitting, we varied model parameters and tried to get the best match between the results of a model and the observed H$\alpha$ line. However our model does not include any effect that enhances the continuum flux from loops \citep{Heinzel2018}, so an additional constant was included to account for the offset.

For the flare in the Fig. \ref{fig:profile_changes_2} the fitted synthetic profile difference is shown in Fig. \ref{fig:observation_model_fit_2}. We fit spectrum observed on 18 April 2019 at 22:07 UT. The fitted parameters are listed in the Table \ref{tab:fitted_model_parameters}. The flare area from the fit covers approximately 15\,\% of the stellar disc surface of AD Leo.
The observed profile difference has a stronger red wing for velocities within 15\,--\,50\,km\,s$^{-1}$. The synthetic profile difference has stronger red wing up to velocities of 100\,km\,s$^{-1}$. Compared to the observed profile difference the synthetic difference has a similar shape but appears to be shifted by a few~km\,s$^{-1}$ towards red velocities.

For the flare in the Fig. \ref{fig:profile_changes_1} the result of our fitting is shown in Fig. \ref{fig:observation_model_fit}. We fit the spectrum observed on 19 April 2019 at 23:42 UT. The fitted parameters are listed in the Table \ref{tab:fitted_model_parameters}. The flare area from the fit covers approximately 16\,\% of the stellar disc of AD Leo. 
The observed profile difference has a stronger red wing for velocities within 15\,--\,50\,km\,s$^{-1}$. The synthetic profile difference has a stronger red wing up to velocities of 100\,km\,s$^{-1}$. Compared to the observed profile difference the synthetic difference has a similar shape but appears to be shifted by a few~km\,s$^{-1}$ towards red velocities as in the previous comparison.

\begin{table*}
    \caption{Fitted model parameter values for the two selected profiles.}
    \label{tab:fitted_model_parameters}
    \centering
    \begin{tabular}{c c c c c c c c c}
    \hline \hline
    Flare & $T$ [K] & $n_e$ [cm$^{-3}$] & $\tau_0$ & $v_{\rm t}$ [km\,s$^{-1}$] & $r$ [Mm] & $A$ [Mm$^2$] & $v_{0}$ [m\,s$^{-1}$] & constant [erg\,Å$^{-1}$\,cm$^{-2}$\,s$^{-1}$] \\
    \hline
       18. 04 . 2019  & 10000 & 10$^{12}$ & 8.2 & 2 & 75 & 16875 & 50 & 7~$\times 10^{-14}$\\
       19. 04 . 2019  & 10000 & 10$^{12}$ & 5.5 & 2 & 80 & 19200 & 50 & 14~$\times 10^{-14}$\\
    \hline
    \end{tabular}
\end{table*}

\begin{figure}
    \centering
    \resizebox{\hsize}{!}{\includegraphics{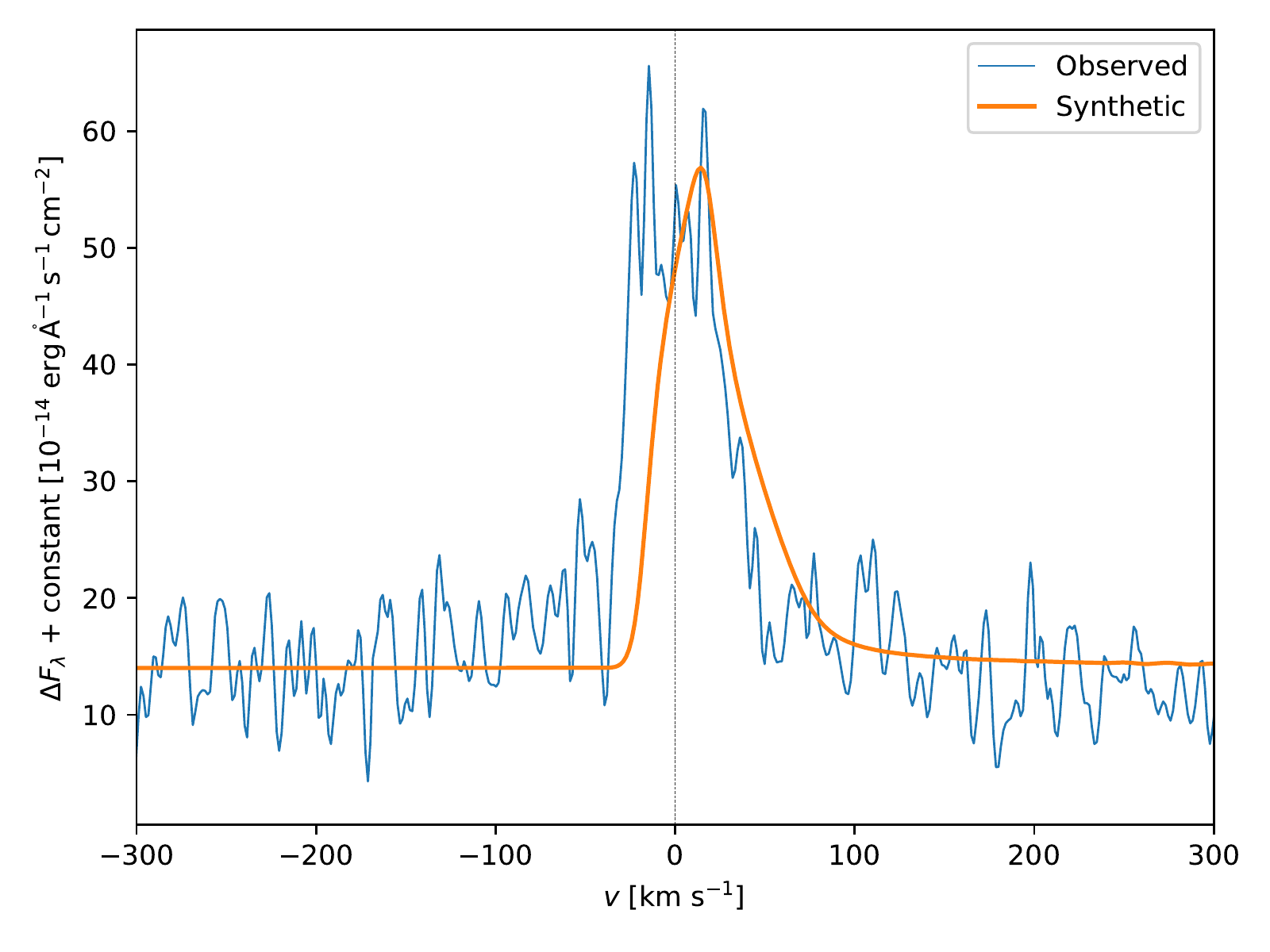}}
    \caption{Fit of the modeled spectrum difference $\Delta F_{\lambda}$ and spectrum difference $\Delta F_{\lambda}$ observed on 19 April 2019 during gradual phase at 23:42 UT in Fig. \ref{fig:profile_changes_1}. To account for the observed continuum enhancements, a constant is added to the synthetic profile difference as a fit parameter. The synthetic profile has a similar shape compared to the observed within the range of 15\,--\,50~km\,s$^{-1}$.}
    \label{fig:observation_model_fit}
\end{figure}

\begin{figure}
    \centering
    \resizebox{\hsize}{!}{\includegraphics{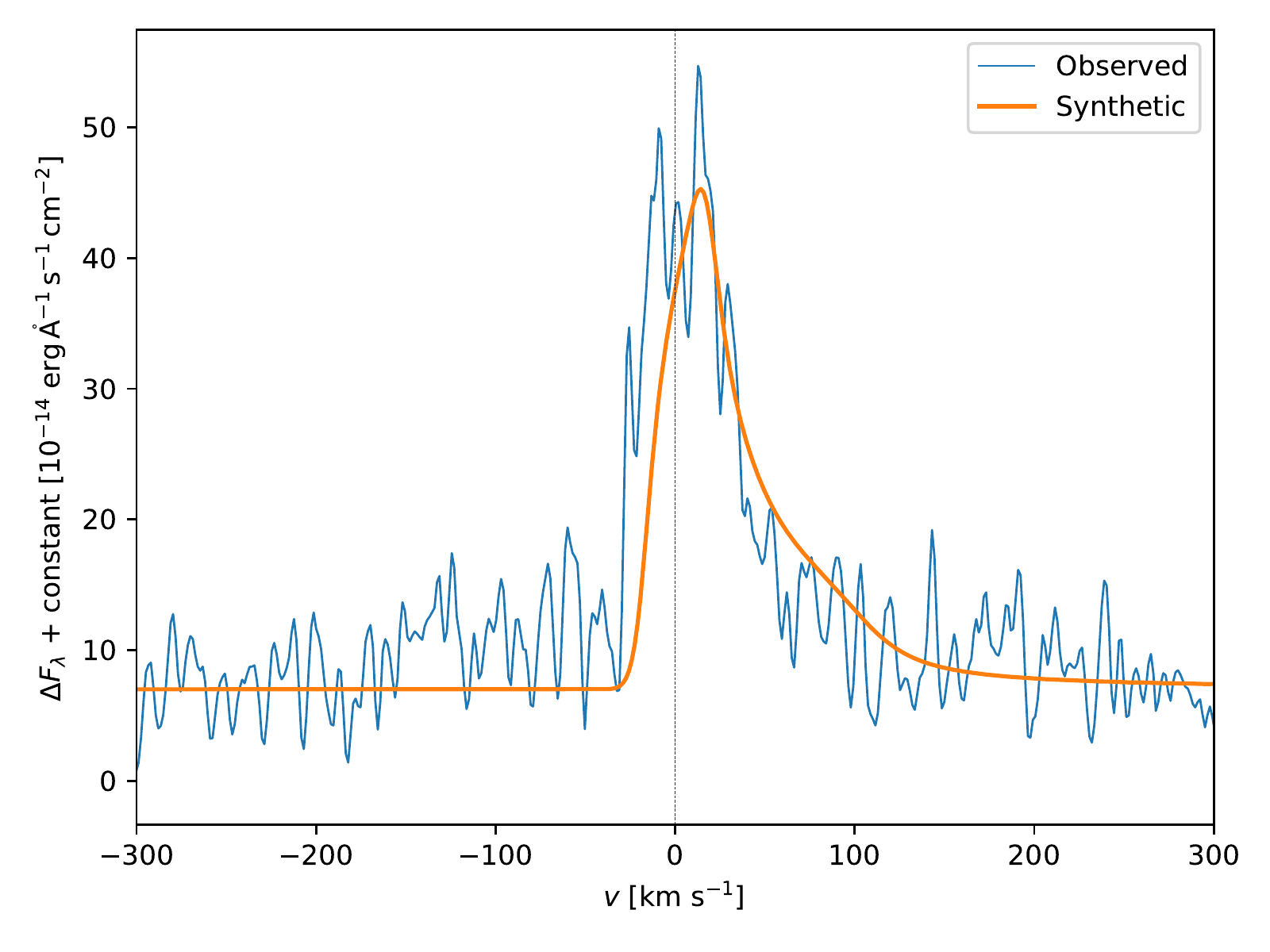}}
    \caption{Fit of the modeled spectrum difference $\Delta F_{\lambda}$ and spectrum difference $\Delta F_{\lambda}$ observed on 18 April 2019 during gradual phase at 22:07 UT in Fig. \ref{fig:profile_changes_2}. To account for the observed continuum enhancements, a constant is added to the synthetic profile difference as a fit parameter. The synthetic profile has a similar shape compared to the observed within the range of 15\,--\,50~km\,s$^{-1}$.}
    \label{fig:observation_model_fit_2}
\end{figure}

\section{Discussion}
Spectral line asymmetries during stellar flares have been observed for several stars in recent years, e.g. \citet{Fuhrmeister2018}, \citet{Vida2019}, \citet{Leitzinger2022}, \citet{Muheki2020}, \citet{Muheki2020cme}, \citet{Koller2021} or \citet{Wu2022}. Enhancements of the blue wing of spectral lines are often linked to the possible presence of CMEs. Enhancements in the red wing are associated with downward flows of the plasma, mostly chromospheric condensation or backward-falling material from an eruption, or CMEs occurring close to the limb. Velocities in the red asymmetry can reach several hundred km\,s$^{-1}$. Our OES observations of the H$\alpha$ line indicate velocities up to 50 km\,s$^{-1}$ while the observed profile changes are similar to asymmetries observed by \citet{Muheki2020}.

On the Sun the observed asymmetries in some spectral lines during impulsive phases of flares are caused by the chromospheric condensation \citep{Kuridze2015} with velocities reaching a few tens of km\,s$^{-1}$. This mechanism could explain some of the lower-velocity red asymmetries observed on M-type stars but the presence of high-velocity red asymmetries (hundreds of km\,s$^{-1}$) suggests other mechanisms to be present. A good candidate seems to be coronal rain, a downflow of plasma usually formed during the gradual phase of flares. It can reach much higher velocities compared to the chromospheric condensation, \citet{Antolin2010} reports velocities up to 120~km\,s$^{-1}$. On cool M-dwarf stars, where flare loops are expected to be much larger compared to the stellar radius than on the Sun, e.g. \citep{Hawley1995}, the velocities of the coronal rain could reach even higher values.

In our model, we solve the H$\alpha$ radiative transfer inside the cool loop clouds. Synthetic profiles resulting from the whole loop arcade are asymmetrical with an enhanced red wing of the line. This enhancement can be followed up to 200~km\,s$^{-1}$ for the parameters we used. Comparing the synthetic profiles of our model with OES observations we are able to reproduce a similar shape of the profile with enhanced red wing suggesting that the presence of a coronal rain could create asymmetries consistent with observations. However, there are three discrepancies that our model doesn't explain.

First, the whole synthetic profile difference appears to be somewhat shifted towards the red by a few km\,s$^{-1}$. That could be caused by a flare occurring not at the center of the stellar disc as our model assumes but rather at a different position. In our model the flare occurs at the center of the stellar disc with respect to the observer. In this case, all of the clouds in the filled arcade of flare loops move away from the observer. If the flare occurred a little further from the center of the disc, one half of the clouds close to the top of the arcade would move towards the observer with lower velocities and the other half of the clouds would move away. The radiation coming from the clouds close to the top moving toward the observer would therefore be blue-shifted rather than red-shifted. Additionally, some of the clouds close to the farther anchor of the flare loops would be covered by the rest of the arcade and the line-of-sight component of terminal velocity of the clouds at the other anchor would be lower. This would result in a slightly stronger blue wing and a slightly weaker red wing virtually shifting the whole profile difference towards blue velocities. If the flare occurred at the limb most of the clouds would contribute radiation in the blue wing rather than the red one possibly creating a blue asymmetry. To account for a general location of the flare will require a more complex approach and this will be a subject of our future studies.

Second, our model does not provide any continuum enhancements that are observed by OES. As discussed in the section \ref{sec:observations}, the observed continua enhancements are caused by both the flare and by uncertainties introduced by using linear interpolation during our flux calibration process. To account for this we just add a constant to the synthetic profile during fitting. The continuum enhancements are often linked to flare ribbons which are bright long and narrow areas. However, \citet{Heinzel2018} showed that on cool stars the flare loops can significantly contribute to the total white-light flux. Unlike on the Sun, we are unable to resolve these features on stars. The observed spectra likely contain contributions of both loop and ribbon components but here we assume that during the gradual phase the loop arcade dominates. It is well known that in later phases of solar flares the ribbons fade out while the loops grow and fill a larger and larger area, for example \citet{Jing2016}.

Third, the profile differences of the H$\alpha$ line during the second flare in the Fig. \ref{fig:profile_changes_1} have a reversal in their center forming a double peak. In our model, we assume a constant H$\alpha$ source function in the cloud which leads, by definition, to a non-reversed profile. The reverted profile can be modeled using the full non-LTE radiative transfer but that is beyond the scope of this paper and could be a subject of future studies.

\section{Conclusions}
We have observed dMe star AD Leo during the spring periods in 2019, 2020, and 2021, using the Ondřejov Échelle Spectrograph (OES) attached to the 2-meter Perek telescope at Ondřejov observatory. Simultaneously, in 2019 and 2021 AD Leo photometric observations were carried out. In this paper, we have studied the effect of stellar flares on the H$\alpha$ line. During flares, we observed that the H$\alpha$ line exhibited enhancement in the line center, broadening, and an asymmetry with enhanced flux in the red wing at velocities within up to 50~km\,s$^{-1}$.

In order to explain these H$\alpha$ profile asymmetries, we developed a simple model based on a
direct analogy with solar flares. The model synthesizes H$\alpha$ profiles emergent from
an arcade of flare loops, assuming that the flare occurs at the center of the stellar disc. The resulting H$\alpha$ profiles are asymmetrical with the enhanced red wing at velocities reaching up to 200~km\,s$^{-1}$.

We attempted to fit the model results to match the observed asymmetries. Our model yields profiles with a similar shape but the whole profile differences appear to be slightly shifted towards the red by a few km\,s$^{-1}$, moreover, our model does not produce any continuum enhancements that OES observed. The whole profile shifting is probably caused by a flare that does not occur at the center of the stellar disc and thus some contribution from the loop arcade is not red-shifted but can be blue-shifted thus effectively shifting the whole profile difference towards the blue wing. More complex geometrical models (projections) must be used to solve this.

\begin{acknowledgements}
This study was supported by the Czech Funding Agency grants GACR 19-17102S and 22-30516K, 
LTT-20015 for data collection with 2-m telescope in 2020 and by RVO:67985815 project.
This work was also partially supported by the program "Excellence Initiative 
- Research University" for years 2020-2026 at University of Wroc{\l}aw, 
project no. BPIDUB.4610.96.2021.KG. This paper is based on the results of the diploma thesis of J. Wollmann.
We would like to thank the observers from the SPHE section of the Czech Astronomical Society for providing us with their photometric observations of AD Leo and namely to H. Kučáková for her photometric observations using the Mayer's telescope at Ondřejov observatory.
We are grateful to M. Špoková, R. Karjalainen, J. Šubjak, and M. Skarka for the reduction of spectra from 2-meter Perek telescope and their advice during further processing of spectra and to E. Guenther for his advice during the spectra calibration process.
We are also grateful to the referee for his/her useful comments.
\end{acknowledgements}

\bibliographystyle{bibtex/aa}
\bibliography{bibtex/references}
\end{document}